\documentclass[journal]{IEEEtran}
\pdfoutput=1
\usepackage{amsmath}
\usepackage{amsfonts}
\usepackage{latexsym}
\usepackage{amssymb}
\usepackage{mathabx}
\usepackage{enumitem,kantlipsum}
\usepackage{epsfig}

\usepackage{hyperref}
\usepackage{bm}
\usepackage{upref}
\usepackage{theorem}
\usepackage{thmtools}
\usepackage{graphicx}
\usepackage{psfrag}
\usepackage{cite}
\usepackage{mathtools}
\mathtoolsset{showonlyrefs}

\allowdisplaybreaks[3]

\usepackage{color}

\usepackage{tikz}
\usetikzlibrary{arrows.meta,bending}

\makeatletter
\newcommand\curvearrowed@[3]
  {%
    \begin{tikzpicture}
      \node[inner sep=.05ex](a){\kern-.25ex#3\kern-.05ex};
      \draw[arrows={-Latex[#1]},line width=#2]
        (a.north west) to[bend left=45] (a.north east);
    \end{tikzpicture}%
  }
\newcommand\curvearrowed[1]
  {%
    \relax
    \ifmmode
      \mathchoice
        {\curvearrowed@{}{.4pt}{$\displaystyle #1$}}
        {\curvearrowed@{}{.4pt}{$\textstyle #1$}}
        {\curvearrowed@{scale=.8}{.325pt}{$\scriptstyle #1$}}
        {\curvearrowed@{scale=.7}{.25pt}{$\scriptscriptstyle #1$}}%
    \else
      \curvearrowed@{}{.4pt}{#1}%
    \fi
  }
\makeatother

\usepackage{algorithm,rotating}
\usepackage{algcompatible} 
\algblockdefx{PARFOR}{ENDPARFOR}[1]%
  {\textbf{for all }#1 \textbf{do in parallel}}%
  {\textbf{end for}}
\makeatletter
\newcommand{\removelatexerror}{\let\@latex@error\@gobble}
\makeatletter
\newcommand{\proofpart}[2]{%
	\par
	\addvspace{\medskipamount}%
	\noindent\emph{Part #1: #2}\par\nobreak
	\addvspace{\smallskipamount}%
	\@afterheading
}
\makeatother

\theoremstyle{plain}
\theorembodyfont{\normalfont\slshape}

\newtheorem{theorem}{Theorem$\!$}
\newtheorem{lemma}{Lemma$\!$}

\newtheorem{corollary}{Corollary$\!$}
\newtheorem{proposition}{Proposition$\!$}

\newtheorem{definition}{Definition$\!$}

\usepackage{amsmath}
\DeclareMathOperator*{\argmax}{arg\,max}
\DeclareMathOperator*{\argmin}{arg\,min}
\DeclareMathOperator{\sign}{sign}

\newcounter{enumrom}
\renewcommand{\theenumrom}{(\roman{enumrom})}

\makeatletter
\renewcommand{\@endtheorem}{\endtrivlist}
\makeatother

\makeatletter
\renewcommand{\thefigure}{{\@arabic\c@figure}}
\renewcommand{\fnum@figure}{{\bf Figure\,\thefigure}}
\makeatother

\newcommand{\cB}{\mathcal{B}}

\newcommand{\cD}{\mathcal{D}}

\newcommand{\cZ}{\mathcal{Z}}

\newcommand{\bc}{\mathbf{c}}

\newcommand{\bI}{\mathbf{I}}

\renewcommand{\le}{\leqslant}
\renewcommand{\leq}{\leqslant}
\renewcommand{\ge}{\geqslant}
\renewcommand{\geq}{\geqslant}

\newcommand{\Cref}[1]{Co\-ro\-lla\-ry\,\ref{#1}}

\newcommand{\dgd}{{\sc DGD}}
\newcommand{\rdgd}{{\sc RDGD}}
\newcommand{\rdgdRS}{{\sc RDGD-Restart}}
\newcommand{\rdgdSC}{{\sc RDGD-SC}}

\outer\def\proclaim #1. #2\par{\medbreak
 \noindent{\bf#1.\enspace}{\sl#2\par}%
 \ifdim\lastskip<\medskipamount \removelastskip\penalty55\medskip\fi}

\begin{document}

% paper title
\title{Robust Distributed Gradient Descent to Corruption over Noisy Channels}
%\title{Distributed Gradient Descent to Robust  Corruption over Noisy Channels}

\title{A Mirror Descent-Based Algorithm for Corruption-Tolerant Distributed Gradient Descent }

\author{Shuche Wang,~\IEEEmembership{Graduate~Student~Member,~IEEE} and
Vincent Y.~F.~Tan,~\IEEEmembership{Senior~Member,~IEEE}

\thanks{S.~Wang is with the Institute of Operations Research and Analytics, National University of Singapore (Email: shuche.wang@u.nus.edu). V.~Y.~F.~Tan is  with the Department of Mathematics, the Department of Electrical and Computer Engineering, and the  Institute of Operations Research and Analytics, National University of Singapore (Email: vtan@nus.edu.sg).}
\thanks{ 
This paper was presented in part at the 2024 IEEE International Symposium on Information Theory (ISIT)~\cite{wang2024robust}.}
\thanks{This work is funded by a Singapore Ministry of Education AcRF Tier~2
grant (A-8000423-00-00) and two AcRF Tier~1 grants (A-8000980-00-00 and A-8002934-00-00).
}
}

\maketitle

\begin{abstract}
Distributed gradient descent algorithms have come to the fore in modern machine learning, especially in parallelizing the handling of large datasets that are distributed across several workers. However, scant attention has been paid to analyzing the behavior of distributed gradient descent algorithms in the presence of adversarial corruptions instead of random noise. In this paper, we formulate a novel problem in which adversarial corruptions are present in a distributed learning system. We show how to use ideas from (lazy) mirror descent to design a corruption-tolerant distributed optimization algorithm.  Extensive convergence analysis for (strongly) convex loss functions is provided for different choices of the stepsize. We carefully optimize the stepsize schedule to accelerate the convergence of the algorithm, while at the same time amortizing the effect of the corruption over time. Experiments based on linear regression, support vector classification, and softmax classification on the MNIST dataset corroborate our theoretical findings. 
\end{abstract}
\begin{IEEEkeywords}
Distributed gradient descent, Mirror descent,  Adversarial corruptions, Noisy channels, Convergence rates
\end{IEEEkeywords}

\section{Introduction}

 Driven by the growing scale of datasets in various learning tasks, distributed gradient descent has become increasingly important in modern machine learning~ \cite{boyd2011distributed,jordan2018communication,low2012distributed}. In this paradigm, there is a central parameter server coordinating information among all workers and each worker in the system processes partial gradients using its subset of data~\cite{richtarik2016distributed,richtarik2016parallel}. These partial gradients are then collectively used to update the global model parameters. Such a distributed framework not only expedites the learning process but also manages datasets that are too large to be handled in the memory of a single worker or server. 

However, there are significant challenges in the use of  distributed gradient descent such as the communication overhead~\cite{chen2018lag}, the necessity of having resilience against system failures~\cite{tandon2017gradient,bitar2020stochastic}, and the presence of  adversarial attacks~\cite{chen2018draco,chen2017distributed,yin2018byzantine,cao2019distributed}. In a distributed framework, workers may become unreliable (called {\em Byzantine} workers in the sequel), which can disrupt the learning process. For instance, as demonstrated in~\cite{blanchard2017machine}, the presence of even a single Byzantine worker can undermine the convergence of a distributed gradient descent algorithm.

In this paper, we consider a new problem concerning the distributed gradient descent algorithm but we take into account a novel and critical aspect that has, to the best of our knowledge, not been considered before. We consider the presence of  adversarial corruptions at each worker in addition to the usual random (Gaussian)  noises in the uplink and downlink channels. The corruptions and noises affect the accuracy of partial gradients and parameters.  Each worker experiences adversarial corruption, with the total corruption budget being bounded over the entire time horizon. An application of our work could be in a   wireless setting with analog transmission of the gradients. However, instead of assuming that this wireless setting is corruption-free, we assume that there are malicious parties in the system that systematically and deterministically modify the gradients that are sent to a central server for downstream machine learning tasks. Our convergence analysis of a modified version of a (lazy) mirror descent-type algorithm addresses the \textit{worst-case} scenario of these corruptions subject to a long-term constraint on the total corruption budget. We  show both theoretically and empirically that our algorithm \rdgd \ and its variants  such as \rdgdRS\  can be suitably tuned to be corruption tolerant. %Besides, we also consider the stochastic noises in both downlink and uplink channels during the transmission of the parameter.

\subsection{Main Contributions}
Our main contributions are as follows:
\begin{enumerate}[leftmargin=*]
    \item We formulate the problem of distributed gradient descent with   adversarial corruptions on the partial gradient vectors held by  each of the $m$ workers. We impose   a bound $C(T)$ on the total amount of corruptions over the $m$ workers and over the time horizon $T$. The uplink and downlink channels are also subject to the usual Gaussian noises. 
    \item We design a mirror descent-inspired distributed gradient descent algorithm \rdgd\ that is robust to corruptions.  For smooth and convex functions, we show that the  expected suboptimality gap of \rdgd\  decays as $O\big(\frac{1}{\sqrt{T}}+ \frac{C(T)}{m\sqrt{T}}\big)$. 
    \item We extend \rdgd\ to the case of loss functions that are  strongly convex in addition to being smooth. We bound the  suboptimality gap of the enhanced algorithm \rdgdSC. The   convergence behavior of \rdgdSC\ depends on different choices of the stepsize.  For a certain choice of stepsize, we observe that the expected suboptimality gap decays as $O\big( \rho^t + \frac{C(t)}{m}\big)$ where $\rho\in(0,1)$ is a contraction factor and $t$ is number of iterations, while for another choice, the gap decays as $O\big( \frac{1}{t^2} + \frac{C(t)}{m\sqrt{t}}\big)$. Thus for the former, there is an exponentially decaying term and a term that does not cause  the corruption to vanish; for the latter, there is a  (slower) polynomially decaying term and a term that amortizes the  corruption. 
    \item  The above observation motivates us to design a {\em hybrid} algorithm  \rdgdRS\  that   exploits the benefits of each choice of the stepsize. \rdgdRS \ switches the stepsize schedule at an analytically determined transition time.   In \rdgdRS, we   first exploit the exponential decrease of the  suboptimality gap, and eventually amortize the corruption by  changing the stepsize. 
    \item The proposed algorithms are assessed numerically through several canonical machine learning models such as  linear regression, support vector classification, and softmax classification (on the MNIST dataset). All experimental results corroborate our theoretical findings.

\end{enumerate}

\subsection{Related Works}
There are many studies that analyze the effect of {\em  random} corruptions in (stochastic) gradient descent~\cite{lan2012optimal,ghadimi2012optimal,devolder2014first} and its accelerated versions~\cite{lan2012optimal,cohen2018acceleration,zhou2022acceleration}. Our study adopts a distinctly different approach in which we focus on the effect of {\em malicious adversarial} corruptions, instead of random ones, on distributed gradient descent-type algorithms. In this scenario, some workers send maliciously corrupted gradient information to adversely affect the system's performance. In~\cite{chang2022gradient}, the authors investigate the robustness of gradient descent in the face of the worst case of corruption, where adversaries can manipulate the corruption up to a certain bounded extent throughout the gradient descent process. In contrast to our {\em  distributed} gradient descent formulation, \cite{chang2022gradient} considers a {\em  centralized} gradient descent scheme. Furthermore, the approach of~\cite{chang2022gradient} does not modify the traditional gradient descent algorithm to further mitigate the adversarial corruption's effect. %Besides the adversarial corruption, it is natural to consider the imperfect communication, where both the upload and download transmissions taking place over noisy channels.

Another line of research that  relevant to our work pertains to  \emph{Byzantine distributed gradient descent}. Byzantine faults refer to the scenario where some nodes in the distributed network act arbitrarily or maliciously, potentially leading to incorrect or suboptimal  outcomes. One line of solutions is based on the {\em robust aggregation rules}, including trimmed mean~\cite{yin2018byzantine}, coordinate-wise median~\cite{yin2018byzantine}, geometric median~\cite{chen2017distributed}, and Krum~\cite{blanchard2017machine}. However, these robust aggregation rules work under the assumption that the number of Byzantine workers is  less than half the total number of workers. Our research approach bifurcates from this line of research since we allow the adversary to {\em arbitrarily} allocate a bounded amount of  corruptions among \textit{all workers}. Our model, however, naturally precludes the possibility of tolerating arbitrary perturbations as median-based approaches that eliminate corruptions of arbitrarily large magnitudes cannot be directly employed.

In addition, these robust aggregation methods can,  in general, only recover the parameters obtained from vanilla gradient descent with high probability but not with probability one. %can only approximately mimic gradient descent. 
To address this challenge,  an algorithm called {\sc DRACO} is presented in \cite{chen2018draco}. {\sc DRACO}  is a framework that applies {\em repetition codes} to the computational results of the  workers. This requires roughly doubling the replication of each gradient computation, i.e., $2s+1$ replications of each partial gradient for $s$ malicious workers. Recently, \cite{hofmeister2023trading} reduced the number of replications from $2s+1$ to $s+1$ by introducing an additional step to first {\em detect} the Byzantine workers. However, if the channels used for  transmitting the parameters are noisy, this idea of replicating each partial gradient does not work perfectly well since the parameter server may fail   to  reliably  select the majority among all the corrupted partial gradients.

The works on adversarial  optimization are also closely related to some lines of work in the  multi-armed bandit literature, particularly in the adversarial (and not stochastic) setting~\cite{ito2021optimal, zimmert2019optimal, zhong2021probabilistic,sachs2022between}. In these settings, adversarial corruption is added to the random rewards, resulting in the agent receiving corrupted versions of the rewards. These approaches typically assume the adversary is capable of arbitrarily allocating corruption in a malicious manner within a total corruption budget. They then use various randomized policies to mitigate the effect of the corruption. However, our study focuses on {\em distributed gradient descent}, which is a first-order optimization method that is rather different scenarios with bandit feedback.

\subsection{Outline}
The rest of this paper is organized as follows. The  framework of distributed
gradient descent in the presence of adversarial corruptions over noisy channels is presented precisely in Section~\ref{sec:problem_setting}. The proposed \rdgd\ algorithm and the analysis of its expected suboptimality gap for smooth loss functions are developed in Section~\ref{sec:smooth}. In Section~\ref{sec:strong}, we additionally assume  the loss function is strongly convex, design an extended algorithm \rdgdSC, and prove a bound on its expected suboptimality gap. The algorithm \rdgdRS\ to mitigate corruption accumulation through a judicious choice of stepsizes is presented in Section~\ref{sec:cor_reduction}. Numerical experiments are presented in Section~\ref{sec:experiments}. Finally, Section~\ref{sec:conclusion} concludes the paper by suggesting several avenues for further research.

\section{Problem Setting}\label{sec:problem_setting}
In this section, we present the framework of distributed gradient descent in the presence of corruptions and noise in channels in both the downlink and the uplink. In addition, we present some basic notations and preliminaries that will facilitate the subsequent analysis.

\subsection{Distributed Gradient Descent}
We assume that the samples  in a dataset $\cZ=\{(x_i,y_i)\}_{i=1}^N$ are drawn from an unknown distribution $\cD$. Here $x_i\in\mathbb{R}^p$ is the $i$-th feature vector and $y_i\in\mathbb{R}$ is the corresponding scalar response (which could be discrete or continuous). Let $\Theta\subset\mathbb{R}^p$ be the set of model parameters. We seek to minimize the average of  loss functions $L:\mathbb{R}^p\times \mathbb{R}\times\Theta\rightarrow \mathbb{R}$  each evaluated over a sample in $\cZ$, i.e., we consider the problem
\begin{equation*}
	\min_{\theta\in\Theta} \Big\{ L(\theta) \coloneq   \frac{1}{N}\sum_{i=1}^N L(x_i,y_i;\theta) \Big\}.
\end{equation*}
%We define $L(\theta)\coloneq\frac{1}{N}\sum_{i=1}^N L(x_i,y_i;\theta).$
In a {\em centralized} gradient descent scheme, the model parameter $\theta_t\in\Theta$ is updated at iteration $t$ as follows, 
\begin{equation*}
\theta_{t+1}=\theta_t-\eta_t g_t,
\end{equation*}
where $g_t:=\frac{1}{N}\sum_{i=1}^N\nabla L(x_i,y_i;\theta_t)$ is the {\em full gradient}   ($\nabla L(x_i, y_i; \theta)$ is the gradient  vector of  $L(x_i, y_i; \theta)$ w.r.t.~$\theta$) and $\eta_t$ is an appropriately chosen stepsize  at iteration $t$.

When $N$ is large, it is inefficient to compute the full gradient $g_t$. Rather, one typically  parallelizes the computation of $g_t$ by distributing the data points among $m\ll N$ workers. We assume the $N$ data samples in $\cZ$ are uniformly partitioned among the $m$ workers.\footnote{Without loss of generality, we assume $N$ is a multiple of $m$.} In particular, the $i$-th worker maintains a subset $\cZ_i$ of the dataset $\cZ$, where $\cZ_i\cap \cZ_j=\emptyset$, $|\cZ_i|=N/m$. Then, worker $i$ computes its {\em partial gradient vector } 
\begin{equation*}
    g_{i,t}:= \frac{1}{|\cZ_i|}\sum_{(x, y) \in \cZ_i} \nabla L(x  , y ; \theta_{t}).
\end{equation*}
  This partial gradient vector is then sent to the server.

\subsection{Noisy channel transmission}
At the $t$-th iteration of  distributed gradient descent, the parameter server broadcasts the current model parameter $\theta_{t}\in\Theta$ to all $m$ workers through noisy downlink channels. Worker $i$ receives a noisy version of $\theta_{t}$ as:
\begin{equation}\label{eq:noise_para}
    \theta_{i,t}=\theta_{t}+v_{t}^{(i)},
\end{equation}
where $v_{t}^{(i)}=[v_{t,1}^{(i)},\dotsc,v_{t,p}^{(i)}]^{\top}\in \mathbb{R}^p$ is the $p$-dimensional noise vector at worker $i$ and time $t$. Furthermore, $v_{t}^{(i)}$ is a zero-mean random vector consisting of independent and identically distributed (i.i.d.) elements with variance $\mathbb{E}[(v_{t,k}^{(i)})^2]=\sigma_t^2$. Thus,
\begin{equation*}
 \mathbb{E}[\|v_{t}^{(i)}\|_2^2]=p\sigma_t^2, \quad\forall i\in[m].
\end{equation*}

Now, the {\em noisy partial gradient vector} at worker $i$ based on its local dataset $\cZ_i$ can be written as
\begin{equation}\label{eq:partial_grad}
    g'_{i,t}:= \frac{1}{|\cZ_i|}\sum_{(x,y ) \in \cZ_i} \nabla L(x , y; \theta_{i,t}),
\end{equation}
where $\theta_{i,t}$ is the noisy parameter vector in \eqref{eq:noise_para}.

\subsection{Corrupted gradients}
At iteration $t$, instead of
transmitting the true local gradients, due to Byzantine faults, each worker also suffers an {\em arbitrary corruption} to its partial gradient vector  as follows: %, where $\varepsilon_{i,t}\in \mathbb{R}^p$.
\begin{equation}
\label{liar}
\bar{g}_{i,t} =
g'_{i,t}+\varepsilon_{i,t}.
\end{equation}
Here the corruption $\varepsilon_{i,t}$ is a vector in $\mathbb{R}^p$. Define the $\ell_2$ norm of the total corruption at iteration $t$ as:
\begin{equation}
    c_t=\bigg\|\sum_{i=1}^m\varepsilon_{i,t}\bigg\|_2. \label{eqn:defct}%, \qquad\sum_{t=1}^T  \mathbb{E}[\|c_t\|^2]\le G(T), 
\end{equation}
This quantifies the aggregate corruption in gradients, introduced by the adversary at iteration $t$. We denote the corruption vector as $c= [c_1,c_2,\dotsc,c_T]^\top\in\mathbb{R}^T$ and assume that there is an overall $\ell_2$ bound on it across the entire time horizon. The form of the bound we employ is
\begin{equation}\label{eq:constraint}
    \|c\|_2=\Big(\sum_{t=1}^T c_t^2\Big)^{1/2}\le C(T).  
\end{equation}
We emphasize that terms $\{\varepsilon_{i,t}\}$ and hence, $\{c_t\}$, and $c$ are {\em deterministic} quantities induced by Byzantine faults. These are in contrast to the {\em stochastic noises} in the (downlink and uplink) channels. Our subsequent analysis is thus  applicable to the {\em worst-case} instantiation of these corruptions subject to the constraint in~\eqref{eq:constraint}.

The $i$-th worker sends $\bar{g}_{i,t}$ defined in \eqref{liar} to the parameter server. However, due to the presence of noise in the uplink communication channels, the parameter server receives the following perturbed version of $\bar{g}_{i,t}$ from worker $i$:
\begin{equation}
    \tilde{g}_{i,t}=\bar{g}_{i,t}+w_{t}^{(i)}, \label{eqn:grad_noise}
\end{equation}
where $w_{i,t}^{(i)}=[w_{t,1}^{(i)},\dotsc,w_{t,p}^{(i)}]^{\top}\in \mathbb{R}^p$ is a $p$-dimensional noise vector at worker $i$ and time $t$. Furthermore, $w_{t}^{(i)}$ is a zero-mean random vector consisting of i.i.d.\ elements with variance $\mathbb{E}[(w_{t,k}^{(i)})^2]=\sigma_t^2$. As a result, 
\begin{equation*}
 \mathbb{E}[\|w_{t}^{(i)}\|_2^2]=p\sigma_t^2, \quad\forall\, i\in[m].
\end{equation*}

An illustration of our distributed gradient descent setting with corruptions over noisy channels is presented in Fig.~\ref{fig:ill_distributed}.

\begin{figure}
    \centering
    \resizebox{0.9\linewidth}{!}{ \input{figs/fig} }
    
    \caption{Illustration of the distributed gradient descent setting with corruption over noisy channels. The  variables indicated in \textcolor{blue}{blue} are random Gaussian noises, defined in \eqref{eq:noise_para} and \eqref{eqn:grad_noise}. The variables indicated in \textcolor{red}{red} are deterministic adversarial corruptions specified in \eqref{liar} and subject to the bound in \eqref{eq:constraint}. }\label{fig:ill_distributed}
\end{figure}

\subsection{Preliminaries}
Consider a continuously differentiable function $f: \Theta \to \mathbb{R}$, where $\Theta \subseteq \mathbb{R}^p$ is a closed and convex set. The convexity of $f$ is characterized by the following inequality:
\begin{equation*}\label{eq:def-cvx}
     f(\theta') \geq f(\theta) + \langle \nabla f(\theta), \theta' - \theta \rangle,\quad\forall \, \theta', \theta \in \Theta.
\end{equation*}
%where $\nabla f(\theta)\in\mathbb{R}^p$ represents the gradient vector of $f$ evaluated at $\theta\in\Theta$.

\begin{definition}[Smoothness]
    A differentiable function $f: \Theta \to \mathbb{R}$ is $M$-smooth on $\Theta$ if for all $\theta, \hat{\theta} \in \Theta$,
    \[ f(\hat{\theta}) \leq f(\theta) + \langle \nabla f(\theta), \hat{\theta} - \theta \rangle + \frac{M}{2} \|\hat{\theta} - \theta\|_2^2. \]
\end{definition}

\begin{definition} [Lipschitz Continuity]
A  function $f: \Theta \to \mathbb{R}$ is {\em $K$-Lipschitz continuous on $\Theta$} if for all $\theta, \hat{\theta} \in \Theta$,
    \[ \|f(\hat{\theta})-f(\theta)\|_2\leq K\|\hat{\theta}-\theta\|_2. \]
\end{definition}

\begin{definition}[Strong Convexity]
    A function $f: \Theta \to \mathbb{R}$ is $\mu$-strongly convex on $\Theta$ if for all $\theta, \hat{\theta} \in \Theta$, 
    \[ f(\hat{\theta}) \geq f(\theta) + \langle \nabla f(\theta), \hat{\theta} - \theta \rangle + \frac{\mu}{2} \|\hat{\theta} - \theta\|_2^2. \]
\end{definition}

\begin{definition}[Convex Conjugate]
    The convex conjugate of a function $\psi: \Theta\subset\mathbb{R}^p \to \mathbb{R}$, denoted as $\psi^*:\mathbb{R}^p\rightarrow\mathbb{R}\cup\{\pm\infty\}$ 
    \[ \psi^*(z) = \max_{\theta \in \Theta} \{ \langle z, \theta \rangle - \psi(\theta) \}. \]
\end{definition}

In the following, we assume that whenever we have to compute the convex conjugate $\psi^*$ of a function $\psi: \Theta \to \mathbb{R}$, this can be done efficiently and potentially in closed form.

%% Danskin's Theorem 
The following is a corollary of Danskin's Theorem~\cite{bertsekas2003convex}.
\begin{proposition}
    Let $\psi: \Theta  \to \mathbb{R}$ be a differentiable and strongly convex function. Then, the gradient of its convex conjugate $\psi^*$ at any point $z$ is given by:
    \[ \nabla \psi^*(z) = \argmax_{\theta \in \Theta} \{ \langle z, \theta \rangle - \psi(\theta) \}. \]
\end{proposition}

\begin{definition}[Bregman Divergence]
    Let $\psi: \Theta \to \mathbb{R}$ be a continuously differentiable and strictly convex function called the {\em mirror map}. The Bregman divergence $D_{\psi}: \Theta^2\rightarrow \mathbb{R}$ is defined as:
    \[ D_{\psi}(\theta, \theta') \coloneq \psi(\theta) - \psi(\theta') - \langle \nabla \psi(\theta'), \theta - \theta' \rangle. \]
\end{definition}

The Bregman divergence $D_{\psi}(\theta, \theta')$ quantities the gap between $\psi(\theta)$ and its first-order Taylor approximation at $\theta'$.

\section{Robust distributed gradient descent to corruption over noisy channel}\label{sec:smooth}
In this section, we present \underline{R}obust \underline{D}istributed \underline{G}radient \underline{D}escent (\rdgd), an algorithm  that is robust to corruptions over noisy channels. For the minimization of smooth loss functions, the algorithm is described in Algorithm~\ref{alg:smooth_mirror}. We also provide a bound on the expected suboptimality gap of \rdgd.

\subsection{The \rdgd \  Algorithm}
\begin{algorithm}[t]
  \caption{Distributed Gradient Descent Algorithm  Robust to Corruptions over Noisy Channels (\rdgd)}\label{alg:smooth_mirror}
  \begin{algorithmic}[1]
   \STATE \textbf{Initialization:} Parameter vector $\theta_0 \in \Theta$, dual vector $z_0=\nabla \psi(\theta_0)$ for some $\mu$-strongly convex function $\psi$, algorithm parameters $\eta_t$, and $ T $. $\theta_1=\theta_0$ and $\hat{\theta}_0=\theta_0$.
   \FOR{$t=1,2,\ldots, T$}
   \STATE \textit{\underline{Parameter server}}: Send $\theta_{t}$ to all $m$ workers over the downlink noisy channel.
   \PARFOR{$i\in[m]$}
   \STATE \textit{\underline{Worker $i$}}:
    \STATE Receive $\theta_{i,t}=\theta_{t}+v_t^{(i)}$ and calculate the corrupted local gradient $g'_{i,t}$ in \eqref{eq:partial_grad}.
   \STATE Send $\bar{g}_{i,t} =
    g'_{i,t}+\varepsilon_{i,t}$ to the parameter server via the noisy uplink channel.
   \ENDPARFOR
   \STATE \textit{\underline{Parameter server}}:
   \STATE Receive $\tilde{g}_{i,t}=\bar{g}_{i,t}+w_t^{(i)}$ for all $i\in[m]$.
    \STATE Compute mean gradient $\tilde{g}_{t}\leftarrow \frac{1}{m}\sum_{i=1}^m\tilde{g}_{i,t}$
   \STATE Update model parameters: 
   \STATE $z_{t}=z_{t-1}-\eta_t \tilde{g}_{t}$, i.e., $z_{t}=-\sum_{k=1}^t \eta_k \tilde{g}_{k}+z_0$.
   \STATE $\hat{\theta}_{t}=\frac{H_{t-1}}{H_t}\hat{\theta}_{t-1}+\frac{\eta_t}{H_t}\theta_t$, where $H_t=\sum_{k=1}^t \eta_k$. 
    \STATE $\theta_{t+1}=\argmin_{u\in \Theta}\{ \sum_{k=1}^t \eta_k \langle \tilde{g}_k, u - \theta_k\rangle  + D_{\psi}(u, \theta_0)\}$.
   \ENDFOR
   \STATE \textbf{Output:} The parameter vector $\hat{\theta}_{T}$.
  \end{algorithmic}
\end{algorithm}

\begin{figure}[t]
    \centering
    \resizebox{0.9\linewidth}{!}{ \begin{tikzpicture}
\centering
%\resizebox{1\textwidth}{!}{%
%\begin{circuitikz}
\tikzstyle{every node}=[font=\large]
\draw  (12.75,10.5) ellipse (0cm and 0cm);
\draw [ color={rgb,255:red,48; green,77; blue,192} , fill={rgb,255:red,209; green,208; blue,240}, rotate around={59:(12.75,10.25)}] (12.75,10.25) ellipse (1.25cm and 1.75cm);
\node [font=\LARGE, color={rgb,255:red,48; green,77; blue,192}] at (13.75,10) {$\Theta$};
\node [font=\LARGE] at (12,6.5) {};
\draw [ color={rgb,255:red,64; green,96; blue,191}] (10.25,13.5) .. controls (9.25,10) and (9.25,10) .. (10.5,6.75);
\draw [ color={rgb,255:red,64; green,96; blue,191}] (5.75,13.5) .. controls (7.75,10) and (6.75,10) .. (5.25,6.75);
\node [font=\LARGE, color={rgb,255:red,86; green,97; blue,184}] at (3,7.25) {Dual Space};
\node [font=\LARGE, color={rgb,255:red,95; green,114; blue,211}] at (13,7.25) {Primal Space};
\draw [->, >=Stealth] (11.75,11.25) .. controls (8,12) and (8,12) .. (4,11.25) ;
\node [font=\large] at (8,11.5) {$\nabla$$\psi$};
\node [font=\Large] at (12,11.25) {$\theta_0$};
\node [font=\Large] at (3.75,11.25) {$z_0$};
\draw [->, >=Stealth] (3.75,11) -- (4,9);
\node [font=\Large] at (3.75,8.5) {$z_{t}=-\sum_{k=1}^{t} \eta_{k} \tilde{g}_{k}+z_0$};
\draw [->, >=Stealth] (4,8) .. controls (8.25,6.75) and (8.25,6.5) .. (12.5,9.25) ;
\node [font=\large] at (8,7.5) {$\nabla\psi^*$};
\node [font=\Large] at (12.75,9.5) {$\theta_{t+1}$};
%\end{circuitikz}
%}

\label{fig:my_label}
\end{tikzpicture} }
    \caption{Illustration of primal and dual updates in \rdgd.}\label{fig:ill_mirror}
\end{figure}

At a high level, we utilize a modified lazy mirror descent approach with primal and dual updates. To initialize, we map the initial point $\theta_0$ to $z_0=\nabla \psi(\theta_0)$ in the dual space via the mirror map $\nabla \psi$ for some suitably chosen $\mu$-strongly convex function $\psi$. At the $t$-th iteration of \rdgd, the parameter server receives the corrupted partial gradients $\tilde{g}_{i,t}$ for all $i\in[m]$. Then, the parameter server calculates the mean of these $\tilde{g}_{i,t}$ and denotes it as $\tilde{g}_{t}= \frac{1}{m} \sum_{i=1}^m \tilde{g}_{i,t}$ (Line~11).  

Next (Line~13), the server updates $z_0$ in the dual space to % ztz_t as 
$
z_t=z_0-\sum_{k=1}^t \eta_k \tilde{g}_{k}. 
$
It then maps $z_t$ back to the point $\theta_{t+1}$ in the primal space (Line 15). The illustration of the primal and dual updates in \rdgd\ is presented in Fig.~\ref{fig:ill_mirror}. We let 
\begin{equation}
\nabla\psi^{*}(z_t)=\argmin_{u\in \Theta}\Big\{ \sum_{k=1}^t \eta_k \langle \tilde{g}_k, u - \theta_k\rangle  +\ D_{\psi}(u, \theta_0)\Big\}. \label{eqn:dual_update}
\end{equation}
This process can be considered as a modified version of lazy mirror descent~\cite{nesterov2009primal}. The intuition behind \rdgd\  is that in~\eqref{eqn:dual_update}, we are minimizing an approximation of the loss function, regularized by the Bregman divergence $D_{\psi}(u,\theta_0)$. The mirror  descent procedure operates in a dual space where the effects of corrupted
gradients can be modulated differently than if done in the primal space. This reduces the impact of the  adversarial corruptions
 since the updates are not directly in the parameter space where the adversary might have
more influence.
Regularization makes the optimization process more sensitive to changes in directions that are
more likely to be influenced by Byzantine workers, effectively dampening their impact. By shaping
the update dynamics through regularization, we limit the influence of corrupted gradients
without significantly hindering the convergence speed of legitimate updates.

We denote 
$
\hat{\theta}_{t}=\frac{H_{t-1}}{H_t}\hat{\theta}_{t-1}+\frac{\eta_t}{H_t}\theta_t,
$
 (where $H_t=\sum_{k=1}^t \eta_k$) as the output at iteration $t$ (Line~14). We note that $\hat{\theta}_{t}$ is the sum of the weighted gradients based on the history of the gradients. As the analysis shows, this ensures that the final output of  \rdgd\ $\hat{\theta}_T$ is robust against noisy updates.

The best choice of the mirror map  $\psi$ depends on the specifics of the task. Commonly used mirror maps include:
\begin{enumerate}
    \item $\ell_2$-norm squared $\psi(x) \,\propto\,\|x\|_2^2$: Suitable for general optimization due to the balance between computational efficiency and being amenable to deriving convergence guarantees.
    \item $\ell_1$-norm $\psi(x) =\|x\|_1$: Promotes sparsity in solutions, making it advantageous in high-dimensional settings.
    \item Negative Entropy $\psi(x) =\sum_ix_i\log x_i$: Effective for tasks involving probability distributions or simplex constraints.
\end{enumerate}
Choosing the appropriate mirror map impacts the algorithm's performance and robustness by tailoring the geometry of the  optimization landscape to the task's requirements. In Section~\ref{sec:ls}, we experimentally compare the efficacy of different mirror maps for our specific setting and conclude that the $\ell_2$-norm squared is the most appropriate, at least for the least squares regression task.

\subsection{Convergence Analysis}
We apply the approximate duality gap technique proposed in~\cite{diakonikolas2019approximate} to analyze the convergence of \rdgd. We will show how to construct an upper bound $G_t$ of the suboptimality gap $L(\hat{\theta}_t)-L({\theta^*})$, where $\hat{\theta}_t$ is the output of a first-order method at the $t$-th iteration. Denote $G_t=U_t-L_t$, where $U_t\ge L(\hat{\theta}_t)$ is an upper bound on $L(\hat{\theta}_t)$ and $L_t\le L({\theta^*})$ is a lower bound on $L({\theta^*})$. Define $H_t\coloneq\sum_{k=1}^t \eta_k$. For the lower bound $L_t$, by the convexity of $L$, we have
\begin{equation}\label{eq:lowerbegin}
L(\theta^*) \geq \frac{\sum_{k=1}^t \eta_k L(\theta_k) + \sum_{k=1}^t \eta_k \langle\nabla L(\theta_k), \theta^* - \theta_k\rangle}{H_t}.
\end{equation}
To express the gradient $\nabla L(\theta_k)$ in relation to the corrupted gradients $\{\tilde{g}_{i,k}: i\in[m]\}$, we write $\nabla L(\theta_k)$ as $\nabla L(\theta_k) = \tilde{g}_{k}+\nabla L(\theta_k)- \tilde{g}_{k}$. Also, by adding and subtracting $\frac{1}{H_t} D_{\psi}(\theta^*, \theta_0)$ in \eqref{eq:lowerbegin}, we have
\begin{align}
&H_tL(\theta^*) \geq  \sum_{k=1}^t \eta_k L(\theta_k) + \sum_{k=1}^t \eta_k \langle \tilde{g}_{k}, \theta^* - \theta_k\rangle \nonumber\\*
&-\sum_{k=1}^t \eta_k\langle \tilde{g}_{k}-\nabla L(\theta_k), \theta^* - \theta_k\rangle + D_{\psi}(\theta^*, \theta_0) - D_{\psi}(\theta^*, \theta_0).\nonumber
\end{align}

Next, we can obtain our lower bound by replacing $\theta^*$
by a minimization over $\Theta$:
\begin{align}
&H_tL(\theta^*) \geq  \sum_{k=1}^t \eta_k L(\theta_k) -\sum_{k=1}^t \eta_k\langle \tilde{g}_{k}-\nabla L(\theta_k), \theta^* - \theta_k\rangle   \nonumber\\* 
&- D_{\psi}(\theta^*, \theta_0)+ \min_{u\in\Theta}\Big\{ \sum_{k=1}^t \eta_k \langle \tilde{g}_{k}, u - \theta_k\rangle+D_{\psi}(u, \theta_0)\Big\} .\nonumber
\end{align}
We then define $H_t L_t$ as the lower bound above, i.e., 
\begin{align}
L_t &\coloneq\frac{1}{H_t}\Big\{\sum_{k=1}^t \eta_k L(\theta_k) -\sum_{k=1}^t \eta_k\langle \tilde{g}_{k}-\nabla L(\theta_k), \theta^*
 - \theta_k\rangle\nonumber\\*
 &\hspace{3cm}- D_{\psi}(\theta^*, \theta_0)+\min_{u\in\Theta} h_t(u)\Big\},\nonumber
\end{align}
where 
\begin{equation}
h_t(u)\coloneq\sum_{k=1}^t \eta_k \langle \tilde{g}_{k}, u - \theta_k\rangle+D_{\psi}(u, \theta_0). \label{eqn:def_ht}
\end{equation}

The promised upper bound is chosen as $U_t = L(\hat{\theta}_{t})$, where $\hat{\theta}_{t}$ is a weighted average of the points $(\theta_k)_{k=1}^t$ generated from \rdgd\, up to $t$-th iteration (specified in Line~14). Given the forms of $U_t$ and $L_t$, it is evident that $L(\hat{\theta}_{t}) - L(\theta^*) \leq G_t$. Therefore, demonstrating the convergence of the algorithm   is tantamount to establishing a bound for $G_t$. We need to track the change of $E_t = H_t G_t - H_{t-1} G_{t-1}$. To facilitate subsequent analyses, we express $G_t$ as:
$$
G_t = \frac{H_1}{H_t}G_1 + \frac{\sum_{k=2}^t E_k}{H_t}.
$$

%\subsection{Convergence Analysis}
According to     \rdgd,  $\theta_{t+1}=\argmin_{u\in \Theta} h_t(u)$. For our analysis, we first   bound $h_t(\theta_{t+1})-h_{t-1}(\theta_t)$ to track the change of the lower bound $L_t$ in the following lemma. The proof is provided in Appendix~\ref{app:argmin_diff}.
\begin{lemma}\label{lem:argmin_diff}
    For all $t\ge 2$, the following holds:
    $$
    h_t(\theta_{t+1})-h_{t-1}(\theta_t)\ge \eta_t\langle \tilde{g}_t, \theta_{t+1}-\theta_t\rangle+D_{\psi}(\theta_{t+1},\theta_t).
    $$
\end{lemma}

\iffalse
\begin{IEEEproof}
    By the definition of $h_t(\cdot)$, we have \begin{equation*}
        h_t(\theta_{t+1})-h_{t-1}(\theta_{t+1})\ge \eta_t\langle \tilde{g}_t, \theta_{t+1}-\theta_t\rangle.
    \end{equation*}
    
    Then, based on the definition of Bregman divergences, we have 
    \begin{multline*}
    h_{t-1}(\theta_{t+1})-h_{t-1}(\theta_{t})=\langle \nabla h_{t-1}(\theta_{t}),\theta_{t+1}-\theta_{t}\rangle\\+D_{h_{t-1}}(\theta_{t+1},\theta_{t})
    \end{multline*}
    and \begin{equation*}
    D_{h_{t-1}}(\theta_{t+1},\theta_{t})=D_{\psi}(\theta_{t+1},\theta_{t}).
    \end{equation*}
    Besides, since $\theta_{t}=\mathrm{argmin}_{u\in\Theta}h_{t-1}(u)$, we have $\langle \nabla h_{t-1}(\theta_{t}),\theta_{t+1}-\theta_{t}\rangle\ge 0$. Then:
    \begin{align*}
        h_t(\theta_{t+1})-h_{t-1}(\theta_{t})\ge \eta_t\langle \tilde{g}_t, \theta_{t+1}-\theta_t\rangle+D_{\psi}(\theta_{t+1},\theta_{t}).
    \end{align*}
\end{IEEEproof}
\fi

Now, we can derive an upper bound on $H_tG_t-H_{t-1}G_{t-1}$ based on Lemma~\ref{lem:argmin_diff}. The proof is provided in Appendix~\ref{app:smoothchange}.
\begin{lemma}\label{lem:smoothchange}
    For all $t\ge 2$, the following holds: %An upper bound of $H_tG_t-H_{t-1}G_{t-1}$ can be shown as:
    \begin{multline*}
    H_tG_t-H_{t-1}G_{t-1}\le \eta_t\langle\tilde{g}_t-\nabla L(\theta_t),\theta^*-\theta_{t+1}\rangle\\
    -\eta_t\langle \nabla L(\theta_t),\theta_{t+1}-\theta_t\rangle-D_{\psi}(\theta_{t+1},\theta_{t}).
    \end{multline*}
\end{lemma}

Finally, we state the convergence rate of \rdgd.
\begin{theorem}\label{thm:smooth}
    Let $0< K,M < \infty$ be such that for all $(x,y)\in\mathbb{R}^p\times \mathbb{R}$, $\ L(x,y;\cdot)$ are  $M$-smooth and $K$-Lipschitz continuous. Let $\theta_0\in\Theta$ be an arbitrary initial point. Let $\{(\theta_t, z_t, \hat{\theta}_t)\}_{t=1}^T$ evolve according to \rdgd \ (Algorithm~\ref{alg:smooth_mirror}) for some $\mu$-strongly convex function $\psi$, where $R_{\Theta}=\max_{\theta\in\Theta}\|\theta-\theta^*\|_2$. Let $\sigma_k\leq \sigma$   for all $k\ge 1$.  For a fixed number of iterations $T$ and stepsize $\eta_k=\frac{1}{\sqrt{T}}$, we have for all $k\ge 1$:
    \begin{align}
        \mathbb{E}[L(\hat{\theta}_T)-L(\theta^*)]&\le \frac{D_{\psi}(\theta^*,\theta_0)}{\sqrt{T}}+\frac{K^2}{2\mu}\cdot\frac{1}{\sqrt{T}}\nonumber\\*
        &+\frac{R_{\Theta}C(T)}{m}\cdot\frac{1}{\sqrt{T}}+(M+1)\sqrt{p}\sigma R_{\Theta}.\nonumber
    \end{align}
\end{theorem}

%For a fixed number of iterations $T$ and the choice of the learning rate $\eta_k=\frac{1}{\sqrt{T}},\forall k\ge 1$, 
The first two terms $\frac{D_{\psi}(\theta^*,\theta_0)}{\sqrt{T}}$ and $\frac{K^2}{2\mu}\cdot\frac{1}{\sqrt{T}}$ are similar to what one would expect for mirror descent~\cite{nemirovskij1983problem}. Since we assume that $\Theta$ is bounded,   $R_{\Theta}$ is finite. The effect of the adversarial corruption is manifested in the term $\frac{R_{\Theta}C(T)}{m}\cdot\frac{1}{\sqrt{T}}$, where recall that $C(T)$  (defined in \eqref{eq:constraint}) represents the total corruption among all $m$ workers over  $T$ iterations. Even though we do not have a lower bound on the gap $ \mathbb{E}[L(\hat{\theta}_T)-L(\theta^*)]$, the scaling  $O \big( \frac{C(T)}{m\sqrt{T}} \big)$ is intuitively order-optimal as, based on~\eqref{eq:constraint}, if each worker suffers a constant amount of corruption at each time, then $C(T) = \Theta (m\sqrt{T})$. Besides, if the variances of the noises in both downlink and uplink channels are bounded across all iterations (i.e., $\sigma_t^2\le \bar{\sigma}^2$ for all $t$ and some $\bar{\sigma}^2<\infty$), the effect of noisy channels on the bound is  $(M+1)\sqrt{p}\sigma R_{\Theta}$. The proof of Theorem~\ref{thm:smooth} is provided in Appendix~\ref{app:smoothmain}.

If the number of iterations $T$ is not known  {\em a priori}, then $\eta_k$ cannot depend on $T$. In this case, we can set $\eta_k=\frac{1}{\sqrt{k}}$ for all $k\ge 1$. A simple corollary of Theorem \ref{thm:smooth} says that under the same assumptions, for all $t\ge1$,
    \begin{align}
        \mathbb{E}[L(\hat{\theta}_t)-L(\theta^*)]&\le \frac{D_{\psi}(\theta^*,\theta_0)}{\sqrt{t}}+\frac{K^2}{2\mu}\cdot\frac{\log t}{\sqrt{t}}\nonumber\\*
        &+\frac{R_{\Theta}C(t)}{m}\cdot\sqrt{\frac{\log t}{t}}+(M+1)\sqrt{p}\sigma R_{\Theta}.\nonumber
    \end{align}
The only difference vis-\`a-vis the convergence rate specified in Theorem~\ref{thm:smooth} is the corruption term $O\big(\sqrt{\frac{\log t}{t}}\big)$, which is slightly worse, by a factor of $\sqrt{\log t}$, than its counterpart in Theorem~\ref{thm:smooth}. This is due to our incognizance of the number of iterations $T$. 

\section{The Strongly Convex Case}\label{sec:strong}
In this section, we make a further assumption that the loss function $L$ is strongly convex (in addition to being smooth), and consequently, we are able to improve on the convergence rate in Theorem~\ref{thm:smooth}. Our extended algorithm   is known as    \rdgdSC \ (Algorithm~\ref{alg:strong}), where the suffix SC stands for \underline{S}trongly \underline{C}onvex. In essence, assuming that the loss function $L$ is $\alpha$-strongly convex, the only step that differs from Algorithm~\ref{alg:smooth_mirror}  is Line~15 therein, which now reads:
\begin{align}
\theta_{t+1}&=\argmin_{u\in \Theta}\Big\{ \sum_{k=1}^t \eta_k \left(\langle \tilde{g}_t, u - \theta_k\rangle + \frac{\alpha}{2}\|u - \theta_k\|^2\right)\nonumber\\* 
& \hspace{2cm}+ D_{\psi}(u, \theta_0)\Big\}.\label{eq:updated_algo}
\end{align}
The purpose of the   quadratic term $\frac{\alpha}{2}\|u-\theta_k\|^2$ in~\eqref{eq:updated_algo} is to exploit the strong convexity of the loss function $L$ to ensure that the expected suboptimality gap decays faster; typically exponentially fast in $t$. 

\begin{algorithm}[t]
  \caption{Distributed Gradient Descent Algorithm  Robust to Corruptions over Noisy Channels for Strongly-Convex Functions (\rdgdSC)}\label{alg:strong}
  \begin{algorithmic}[1]
   \STATE \textbf{Initialization:} Parameter vector $\theta_0\in\Theta$, algorithm parameters $\eta_t$, dual vector $z_0=\nabla \psi(\theta_0)$ with $\psi(x)=\frac{\eta_1\alpha} {2}\|x\|_2^2$,  $\theta_1=\theta_0$ and $\hat{\theta}_0=\theta_0$.
   \FOR{$t=1,2,\ldots$}
   \STATE \textit{\underline{Parameter server}}: Send $\theta_{t}$ to all the workers over the downlink noisy channel.
   \PARFOR{$i\in[m]$}
   \STATE \textit{\underline{Worker $i$}}:
    \STATE Receive $\theta_{i,t}=\theta_{t}+v_t^{(i)}$ and calculate the corrupted local gradient $g'_{i,t}$ in \eqref{eq:partial_grad}.
   \STATE Send $\bar{g}_{i,t} =
    g'_{i,t}+\varepsilon_{i,t}$ to parameter server via the noisy uplink channel.
   \ENDPARFOR
   \STATE \textit{\underline{Parameter server}}:
   \STATE Receive $\tilde{g}_{i,t}=\bar{g}_{i,t}+w_t^{(i)}$, for all $i\in[m]$.
    \STATE Compute mean gradient $\tilde{g}_{t}\leftarrow \frac{1}{m}\sum_{i=1}^m\tilde{g}_{i,t}$.
   \STATE Update model parameter: 
   \STATE $z_{t}=z_{t-1}-\eta_t \tilde{g}_{t}$, i.e., $z_{t}=-\sum_{k=1}^t \eta_k \tilde{g}_{k}+z_0$.
   \STATE $\hat{\theta}_{t}=\frac{H_{t-1}}{H_t}\hat{\theta}_{t-1}+\frac{\eta_t}{H_t}\theta_t$, where $H_t=\sum_{k=1}^t \eta_k$. 
    \STATE $\theta_{t+1}=\argmin_{u\in \Theta}\{ \sum_{k=1}^t \eta_k (\langle \tilde{g}_t, u - \theta_k\rangle + \frac{\alpha}{2}\|u - \theta_k\|^2) + D_{\psi}(u, \theta_0)\}$.
   \ENDFOR
   \STATE \textbf{Output:} The sequence of parameter vectors $\{\hat{\theta}_{t}\}_{t=1}^\infty$.
  \end{algorithmic}
\end{algorithm}

To analyze the convergence  behavior of \rdgdSC, we also apply the approximate duality gap technique used in Section~\ref{sec:smooth}. For the sake of brevity, we only emphasize the parts that are different from Section~\ref{sec:smooth} in this section. For the lower bound $L_t$, exploiting      strongly convexity instead of regular convexity, we redefine  it as 
\begin{align*}
    L_t \coloneq \frac{1}{H_t}\Big\{\sum_{k=1}^t \eta_k L(\theta_k)& -\sum_{k=1}^t \eta_k\langle \tilde{g}_{k}-\nabla L(\theta_k), \theta^* - \theta_k\rangle \\ &- D_{\psi}(\theta^*, \theta_0)+ \min_{u\in\Theta}h_t(u)\Big\}
\end{align*}
where $h_t$ is also redefined as 
\begin{equation}
    h_t(u)\coloneq\sum_{k=1}^t \eta_k \Big(\langle \tilde{g}_{k}, u - \theta_k\rangle+\frac{\alpha}{2}\|u-\theta_k\|_2^2\Big)+D_{\psi}(u, \theta_0). \label{eqn:def_ht_sc}
\end{equation}
The template for our analysis is similar to that in the smooth case. The following definitions are identical, starting with  
$\theta_{t+1}=\mathrm{argmin}_{u\in \Theta} h_t(u)$ and $G_t=U_t-L_t$. Additionally, we need to track the change of $E_t = H_t G_t - H_{t-1}G_{t-1}$, which leads to $G_t=\frac{H_1}{H_t}G_1+\frac{\sum_{k=2}^t E_k}{H_t}$. The following lemma  bounds  $H_1G_1$ and its proof is provided in Appendix~\ref{app:initial_lower}.
\begin{lemma}\label{lem:initial_lower}
    Let $\psi(x)=\frac{\eta_1\alpha}{2}\|x\|_2^2$, the following holds:
    \begin{equation*}
    H_1G_1\le \frac{\eta_1\alpha}{2}\|\theta^*-\theta_1\|_2^2+\eta_1\langle \tilde{g}_{1}-\nabla L(\theta_1), \theta^*-\theta_2\rangle+\frac{\eta_{1}K^2}{4\alpha}.
     \end{equation*}
\end{lemma}
\iffalse
\begin{IEEEproof}
    The initial lower bound is:
    \begin{multline*}
    H_1L_1=\eta_1L(\theta_1)+\eta_1\langle \tilde{g}_{1}, \theta_2-\theta_1\rangle+(\frac{\eta_1\alpha}{2}+\frac{\alpha_0}{2})\|\theta_2-\theta_1\|_2^2\\-\frac{\mu_0}{2}\|\theta^*-\theta_1\|_2^2-\eta_1\langle \tilde{g}_{1}-\nabla L(\theta_1), \theta^* - \theta_1\rangle.
    \end{multline*}
The upper bound can be written as:
    \begin{align*}   
    H_1U_1=\eta_1L(\hat{\theta}_1)=\eta_1L(\theta_1)
    \end{align*}
Then, we have:
    \begin{align*}
        &H_1U_1-H_1L_1=H_1G_1\\
        &\le\frac{\alpha_0}{2}\|\theta^*-\theta_1\|_2^2+\eta_1\langle \tilde{g}_{1}-\nabla L(\theta_1), \theta^*-\theta_2\rangle\\&\qquad\quad-\eta_1\langle \nabla L(\theta_1), \theta_2-\theta_1\rangle-(\frac{\eta_1\alpha}{2}+\frac{\alpha_0}{2})\|\theta_2-\theta_1\|_2^2\\
        &\le \frac{\alpha_0}{2}\|\theta^*-\theta_1\|_2^2+\eta_1\langle \tilde{g}_{1}-\nabla L(\theta_1), \theta^*-\theta_2\rangle+\frac{\eta_1^2M^2}{2(\eta_1\alpha\!+\!\alpha_0)}\\
        &=\frac{\eta_1\alpha}{2}\|\theta^*-\theta_1\|_2^2+\eta_1\langle \tilde{g}_{1}-\nabla L(\theta_1), \theta^*-\theta_2\rangle+\frac{\eta_1M^2}{4\alpha}
    \end{align*}
\end{IEEEproof}
\fi

The following lemma   bounds $h_t(\theta_{t+1})-h_{t-1}(\theta_{t})$ and serves  to track the change of the lower bound $L_t$. Its proof is provided in Appendix~\ref{app:stronglower}.
\begin{lemma}\label{lem:stronglower}
    Let $\psi(x)=\frac{\eta_1\alpha}{2}\|x\|_2^2$. For all $t\ge 2$, 
    \begin{align}
    h_t(\theta_{t+1})-h_{t-1}(
    \theta_{t}) &\ge \eta_t\langle \tilde{g}_t, \theta_{t+1}-\theta_t\rangle\\
    &\qquad+\frac{H_{t}\alpha}{2}\|\theta_{t+1}-\theta_t\|_2^2.
    \end{align}
\end{lemma}
\iffalse
\begin{IEEEproof}
    By the definition of $h_t(\cdot)$, we have $h_t(\theta_{t+1})-h_{t-1}(\theta_{t+1})\ge \eta_t\langle \tilde{g}_t, \theta_{t+1}-\theta_t\rangle+\frac{\eta_t\alpha}{2}\|\theta_{t+1}-\theta_t\|_2^2$.
Then, based on the definition of Bregman divergences:
    \begin{multline*}
        h_{t-1}(\theta_{t+1})-h_{t-1}(\theta_{t})=\langle \nabla h_{t-1}(\theta_{t}),\theta_{t+1}-\theta_{t}\rangle\\+D_{h_{t-1}}(\theta_{t+1},\theta_{t})
    \end{multline*}
    Since $\theta_{t}=\mathrm{argmin}_{u\in\Theta}h_{t-1}(u)$, we have $\langle \nabla h_{t-1}(\theta_{t}),\theta_{t+1}-\theta_{t}\rangle\ge 0$. Then:
    \begin{align*}
        &h_{t-1}(\theta_{t+1})-h_{t-1}(\theta_{t})\ge D_{h_{t-1}}(\theta_{t+1},\theta_{t})\\
        &=\frac{H_{t-1}\alpha}{2}\|\theta_{t+1}-\theta_{t}\|_2^2+\frac{\eta_1\alpha}{2}\|\theta_{t+1}-\theta_{t}\|_2^2\\
        &= \frac{H_{t}\alpha}{2}\|\theta_{t+1}-\theta_{t}\|_2^2.
    \end{align*}
\end{IEEEproof}
\fi

Finally, the following lemma whose proof is provided in  Appendix~\ref{app:stronggap}  provides a   bound of $H_tG_t-H_{t-1}G_{t-1}$. % and the proof is provided in
\begin{lemma}\label{lem:gap_strong}
    Let $\psi(x)=\frac{\eta_1\alpha}{2}\|x\|_2^2$ %and $0\le \frac{\eta_t}{H_t}\le \frac{\alpha}{M}$. 
    Then, for all $t\ge 2$, 
    \begin{equation*}
    H_tG_t-H_{t-1}G_{t-1}\le \eta_t\langle \tilde{g}_{t}-\nabla L(\theta_t), \theta^* - \theta_{t+1}\rangle+\frac{\eta_t K^2}{2\alpha}.
    \end{equation*}
\end{lemma}

Equipped with these preliminary results, the convergence rate of   \rdgdSC \ can be quantified as follows.
\begin{theorem}\label{thm:strong}
    Let $0< \alpha\le M < \infty$ and $0<K < \infty$ be such that for all $(x,y)\in\mathbb{R}^p\times \mathbb{R}$, $\ L(x,y;\cdot)$ is  $M$-smooth, $K$-Lipschitz continuous and $\alpha$-strongly convex. let $\theta_0\in\Theta$ be an arbitrary initial point. Let $\{(\theta_t,  z_t ,  \hat{\theta}_t)\}_{t=1}^T$ evolve according to \rdgdSC \ (Algorithm~\ref{alg:strong}) for $\psi(x)=\frac{\eta_1\alpha}{2}\|x\|_2^2$. Let $R_{\Theta}=\max_{\theta\in\Theta}\|\theta-\theta^*\|_2$ and $\sigma_k\leq \sigma$ for all  $k\ge 1$. Let $\eta_1=1$ and $\eta_k$ satisfy $\eta_k\le  \frac{\alpha}{M}H_k $ for all $k\ge 2$. %Define $a_k \coloneq \frac{\eta_k}{H_k}$.  
    Then for all $t\ge 2$:
    \begin{align}
        \mathbb{E}[L(\hat{\theta}_t)-L(\theta^*)]&\le \Big(\prod_{k=2}^{t} \big(1-\frac{\eta_k}{H_k}\big) \Big)\frac{\alpha}{2}\|\theta^*-\theta_0\|_2^2+\frac{ K^2}{2\alpha}\nonumber\\*
        &\hspace{-2cm}+\frac{R_{\Theta}C(t)}{m}\cdot\frac{\sqrt{\sum_{k=1}^t \eta_k^2}}{H_t}+(M+1)\sqrt{p}\sigma R_{\Theta}. \label{eqn:smooth_sc}
    \end{align}
\end{theorem}
\begin{IEEEproof}
We write $G_t=\frac{H_1}{H_t}G_1+\frac{\sum_{k=2}^t E_k}{H_t}$ as
\begin{align*}
G_t&=\frac{H_1}{H_2}\frac{H_2}{H_3}\dotsm\frac{H_{t-1}}{H_{t}}G_1+\frac{\sum_{k=2}^tE_k}{H_t}\\
 &=\Big(1-\frac{\eta_2}{H_2}\Big)\Big(1-\frac{\eta_3}{H_3}\Big)\ldots\Big(1-\frac{\eta_t}{H_t}\Big)G_1+\frac{\sum_{k=2}^t E_k}{H_t}.
\end{align*}
The remaining parts of the proof are essentially the same as those for Theorem~\ref{thm:smooth}.
\end{IEEEproof}

Observe that we have  the flexibility of setting different stepsize sequences  $\{\eta_k\}_{k=1}^\infty$  to achieve different convergence rates in~\eqref{eqn:smooth_sc}. %We set specific stepsizes $\{\eta_k\}_{k=1}^\infty$ to 
Motivated by this observation, we derive the following corollary of Theorem~\ref{thm:strong}. The proof is presented in Appendix~\ref{app:strongstep}.
\begin{corollary}(of Theorem~\ref{thm:strong}).\label{cor:strongstep} 
    For $k\ge 2$, setting
    \begin{itemize}
    \item  $ \eta_k = \frac{\alpha}{M}H_k$, we have
    \begin{align}
        \mathbb{E}[L(\hat{\theta}_t)-L(\theta^*)]&\le \Big(1-\frac{\alpha}{M} \Big)^{t-1}\frac{\alpha}{2}\|\theta^*-\theta_0\|_2^2+\frac{ K^2}{2\alpha}\nonumber\\*
        &+\frac{R_{\Theta}C(t)}{m}+(M\!+\!1)\sqrt{p}\sigma R_{\Theta}.  \label{eqn:case1}
    \end{align}
    \item  $\eta_k= \frac{2}{k+1}H_k$,  we have
    \begin{align}
        \mathbb{E}[L(\hat{\theta}_t)-L(\theta^*)]&\le O\Big(\frac{\|\theta^*-\theta_0\|_2^2}{t^2}+ \frac{1}{\sqrt{t}}\cdot\frac{R_{\Theta}C(t)}{m}\Big)\nonumber\\*
        &\quad+\frac{ K^2}{2\alpha}+(M+1)\sqrt{p}\sigma R_{\Theta}. \label{eqn:case2}
    \end{align}
    \end{itemize}   
\end{corollary}

If we set $\eta_k = \frac{\alpha}{M}H_k$ (the first case in Corollary~\ref{cor:strongstep}), we notice that the leading term \( \left( 1 - \frac{\alpha}{M} \right)^{t-1} \frac{\alpha}{2} \|\theta^* - \theta_0\|_2^2 \) resembles the convergence behavior typically observed in the analysis of smooth and strongly convex functions, in scenarios without corruption and noise~\cite{garrigos2023handbook}. This  term   decays exponentially fast as~$t$ grows. However, there is an additional term of the form $\frac{R_{\Theta}C(t)}{m}$ that represents the effect of the corruption and does not vanish with~$t$. Alternatively, if we set $\eta_k=\frac{2}{k+1}H_k$ (the second case of Corollary~\ref{cor:strongstep}), the leading term vanishes at a slower rate of $t^{-2}$ but the corruption term vanishes at a rate $\frac{C(t)}{m \sqrt{t}}$. This naturally begs the question of whether we can design a  ``best of both worlds'' algorithm to exploit the benefits of both settings of the stepsize $\eta_k$   to accelerate the overall convergence.

\section{Corruption Reduction by Exploiting the Best of Both Worlds}\label{sec:cor_reduction}
Based on the results from Section~\ref{sec:strong}, we observe that for certain natural choices of the stepsize such as $\eta_k = \frac{\alpha}{M}H_k$ with $H_k=\sum_{\ell=1}^k \eta_{\ell}$, the adversarial corruption results in the upper bound on the suboptimality gap in \eqref{eqn:case1} stagnating at a positive, non-vanishing constant. This is clearly undesirable.  In this section, we explore how to mitigate this corruption accumulation by proposing an algorithm that switches the stepsize schedule at an appropriate transition time. The main idea is to first exploit the exponential decrease of the expected suboptimality gap by setting $\eta_k = \frac{\alpha}{M}H_k$ until the corruption accumulation begins to dominate the convergence behavior. At this point, we switch to a different stepsize schedule  (i.e., $\eta_k = \frac{2}{k+1}H_k$), allowing the corruption to be gradually amortized  and thus enabling the hybrid algorithm to further reduce the suboptimality gap as $t$ grows.

In this section, for clarity, we specify the total corruption budget as $ C(t) = m   t^r$ for some fixed $r\in(0,1/2)$. We observe  from~\eqref{eqn:case1} that the suboptimality gap initially decreases and then increases due to the accumulation of the corruption \( C(t) \) when we set \( \eta_k = \frac{\alpha}{M}H_k \).
Hence, we   determine the {\em optimal transition time} \( t_0 \)  corresponding to the point where the minimum upper bound of suboptimality gap is achieved. The proof of the following lemma is provided in Appendix~\ref{app:transtime}.

\begin{lemma}\label{lem:trans_time}
If $\eta_k = \frac{\alpha}{M}H_k$, the minimum of the upper bound of $\mathbb{E}[L(\hat{\theta}_t)-L(\theta^*)]$ in~\eqref{eqn:case1} with respect to $t \in \mathbb{N}$ is attained at
\begin{equation}
t_0=\Big\lceil-\frac{(1-r)M}{\alpha}W_{-1}\big(B(r)\big)\Big\rceil, \label{eqn:def_t0}
\end{equation}
where $B(r):=-\frac{\alpha}{(1-r)M}\big(\frac{2Mr}{\alpha^2 R_{\Theta}\exp(\alpha/M)}\big)^{\frac{1}{1-r}}$, and $W_{-1}$ denotes the Lambert W function, i.e., the inverse of  $f(w) = we^w$.
\end{lemma}

\iffalse
From the previous lemma, if our target accuracy $\epsilon>\epsilon_0$, we can set the stepsize $a_k=\frac{\eta_k}{H_k}= \frac{\alpha}{M}$. Let $\epsilon'=\epsilon-(\frac{M^2}{2\alpha}+(M+1)\sqrt{p}\sigma R_{\Theta})$. Then, we have the two conditions ensuring $\mathbb{E}[L(\hat{\theta}_t)-L(\theta^*)]\le \epsilon$:
\begin{equation*}
(1-\frac{\alpha}{M})^{t-1}\frac{\alpha}{2}\|\theta^*-\theta_0\|_2^2\le \frac{\epsilon'}{2}\Longleftrightarrow t\ge \frac{M}{\alpha}\log \frac{\alpha R^2}{\epsilon'}+1,
\end{equation*}
\begin{equation*}
\frac{R_{\Theta}\sqrt{t}}{m}\le \frac{\epsilon'}{2}\Longleftrightarrow t\leq (\frac{m\epsilon'}{2R})^2
\end{equation*}
Therefore for ensuring $\mathbb{E}[L(\hat{\theta}_t)-L(\theta^*)]\le \epsilon$ is to perform $t=\frac{M}{\alpha}\log \frac{\alpha R^2}{\epsilon'}+1$ iterations.
\fi

Note from Lemma~\ref{lem:trans_time} that the  {\em transition time} $t_0$ can  be {\em analytical determined} based on known parameters of the problem such as $M$, $\alpha$, and $R_\Theta$. Inspired by Lemma~\ref{lem:trans_time}, we propose the Algorithm~\ref{alg:restart} that is based on \rdgd \ but with a novel {\em  restart} mechanism (\rdgdRS). \rdgdRS \ exploits  the exponential decrease in the suboptimality gap until the effect of the accumulation of the corruption starts  to dominate the overall bound in \eqref{eqn:case1}. To mitigate this, we then switch to a stepsize sequence that decays more slowly to prevent excessive corruption accumulation. This then leads to the bound in  \eqref{eqn:case2} dominating the overall convergence behavior subsequently.

\begin{algorithm}[t]
  \caption{Distributed Gradient Descent Algorithm  Robust to Corruptions over Noisy Channels with Restart (\rdgdRS)}\label{alg:restart}
  \begin{algorithmic}[1]
   \STATE \textbf{Initialization:} Parameter vector $\theta_0\in\Theta$, $\eta_1=1$, dual vector $z_0=\nabla \psi(\theta_0)$ with $\psi(x)=\frac{\alpha}{2}\|x\|_2^2$,  $\theta_1=\theta_0$ and $\hat{\theta}_0=\theta_0$ (defined in \eqref{eqn:def_t0}).

   \STATE \textbf{Stage 1:} Run \rdgdSC \ (Algorithm \ref{alg:strong}) with $\eta_k = \frac{\alpha}{M}H_k$ from $k \geq 2$. Terminate when $k= t_0$.

   \STATE \textbf{Stage 2:} Restart \rdgdSC \ taking $\theta_{t_0}$ as the initial point and change the stepsize to $\eta_k=\frac{2}{k+1}H_k$ for all $k\ge t_0+1$. 
   \STATE\textbf{Output:} The parameter vector $\hat{\theta}_{t}$.
  \end{algorithmic}
\end{algorithm}

We   now bound the suboptimality gap $\mathbb{E}[L(\hat{\theta}_t)-L(\theta^*)]$ of \rdgdRS. The proof is presented in Appendix~\ref{app:errorrestart}.
\begin{theorem}\label{thm:errorrestart}
Under the same setting as Theorem~\ref{thm:strong} and when corruption budget is paramterized as $C(t)=m t^r$ for $r\in(0,1/2)$, for all $t> t_0$ (defined in \eqref{eqn:def_t0}), the convergence rate of \rdgdRS \ in Algorithm~\ref{alg:restart} is:
\begin{multline*}
\mathbb{E}[L(\hat{\theta}_t)-L(\theta^*)]\leq \frac{\alpha}{2}\Big(1-\frac{\alpha}{M}\Big)^{t_{0}-1}\frac{t_{0}(t_{0}+1)}{t(t+1)}\|\theta^*-\theta_0\|_2^2\\+R_{\Theta}t^r\cdot A(t,t_0)+\frac{K^2}{2\alpha}+(M+1)\sqrt{p}\sigma R_{\Theta},
\end{multline*}
where $A(t,t_0) \coloneq \sqrt{\left(\frac{t_0(t_0+1)}{t(t+1)}\right)^2\!+\!\left(\frac{2}{t(t+1)}\right)^2\sum_{k=t_0+1}^t k^2}$.
\end{theorem}
As $t\to\infty$,  $A(t,t_0)$ decays as $O(1/\sqrt{t})$. We observe from the first term of the bound that in the initial rounds (when $t\le t_0$), the suboptimality gap decays rapidly due to the exponentially decaying term with contraction factor $1-\frac{\alpha}{M}$. When $t>t_0$, we observe from the second term that  the effect of the corruption is amortized because $r<1/2$ and $A(t,t_0)=O(1/\sqrt{t})$. 

\section{Numerical Experiments}\label{sec:experiments}
In this section, we illustrate the  performance of   \rdgd \ and its variants  through several numerical experiments, including least squares regression and support vector machine (SVM) classification on synthetically generated datasets, and softmax classification on the MNIST handwritten digits dataset~\cite{lecun1998gradient}. We compare our proposed algorithms with the  vanilla distributed gradient descent (\dgd) algorithm, i.e., the parameter  vector is simply updated as $\theta_{t+1}=\theta_t-\eta_{t}\tilde{g}_t$ at the $t$-th iteration. Note that \dgd\ does not attempt to correct for the presence of corruptions; hence, we expect that it is not robust. 

\subsection{Least Squares Regression on  Synthetic  Datasets} \label{sec:ls}
We start with least squares regression on synthetic datasets. The datasets  contain $N=10,000$ data samples. In particular, the loss function of the least-squares regression is  as  $L(x,y;\theta)=\frac{1}{2}(x^\top\theta-y)^2$ %$L(x_i,y_i;\theta)=\frac{1}{2N}\sum_{i=1}^N(x_i^\top\theta-y_i)^2$ 
where $\theta\in \mathbb{R}^p$ is the   parameter vector with  dimensionality  $p=20$. Then, the loss function computed over the entire dataset $\cZ$ is $L(\theta)=\frac{1}{2N}\sum_{i=1}^N(x_i^\top\theta-y_i)^2$. We consider one parameter server and $m=20$ workers. 

Each element of the feature vectors $x_i$ is drawn from the standard Gaussian distribution $\mathcal{N}(0,1)$. We also randomly generate each dimension of the unknown  parameter $\theta^*$ from $\mathcal{N}(0,1)$. The scalar responses $y_i$ are obtained by taking the inner product of the feature vectors $x_i$ and the unknown parameter $\theta^*$, followed by adding standard Gaussian noise   $\mathcal{N}(0,1)$. These data samples are then uniformly distributed to all $20$ workers.  We set the variances of the Gaussian noise in both downlink and uplink channels to $\sigma^2=0.5$. 

The performance metric  is the suboptimality gap $\mathbb{E} [ L(\hat{\theta}_t)-L(\theta^*)]$. The results reported are averaged over 100 independent trials. We also report the standard deviations across these 100 experiments. The bound of the total corruption until time $t$ is set to $C(t)=20t^{0.4}$. As it is difficult to find the worst case corruption pattern, we use the following procedure as a proxy to generate the corruption $\varepsilon_{i,t}$ at each worker  $i \in [m]$ and at each time $t$. In particular, let $\Gamma(\tau) = \sum_{k=1}^\tau c(k)^2$ denote the corruption budget used up to time  and including $n$. Then, the available corruption budget at time $t$ is $\sqrt{C(t)^2-\Gamma(t-1)}$. Then, we distribute this available corruption budget to each worker randomly. The sign of adversarial corruption at each worker $\sign(\varepsilon_{i,t})$ is set to $-\sign(g'_{i,t})$ for all $i\in[m]$. Also, the stepsize is set to $\eta_t=1/\sqrt{t}$.

\begin{figure}
    \centering
    \includegraphics[width=9cm]{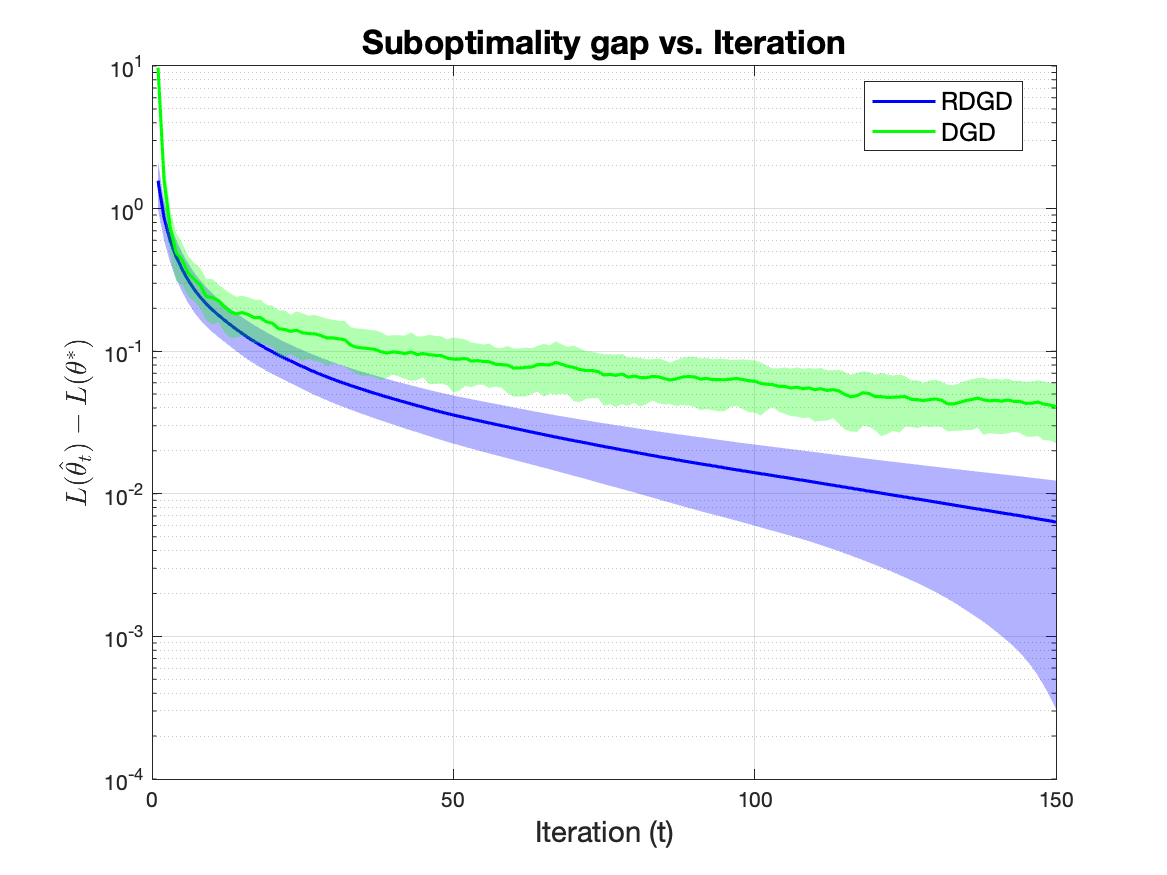}
    \caption{Performances of \rdgd  \  and \dgd \  for least squares regression}\label{fig:linear_smooth}
\end{figure}

Let $X\in\mathbb{R}^{N\times p}$ be the matrix where each row contains the feature vector $x_i^\top$. Define $\lambda_{\mathrm{max}}(Q)$ and $\lambda_{\mathrm{min}}(Q)$ as the maximum and minimum eigenvalues of the matrix $Q$ respectively. We run \rdgd \ (Algorithm~\ref{alg:smooth_mirror}) with the function $\psi(x)=\frac{M}{2}\|x\|_2^2$, where $M=\lambda_{\mathrm{max}}(\frac{1}{N}X^\top  X)$. As shown in Fig.~\ref{fig:linear_smooth}, our proposed algorithm achieves a significantly smaller suboptimality gap more quickly and consistently compared to the traditional \dgd \ algorithm.

\begin{figure}
    \centering
    \includegraphics[width=9cm]{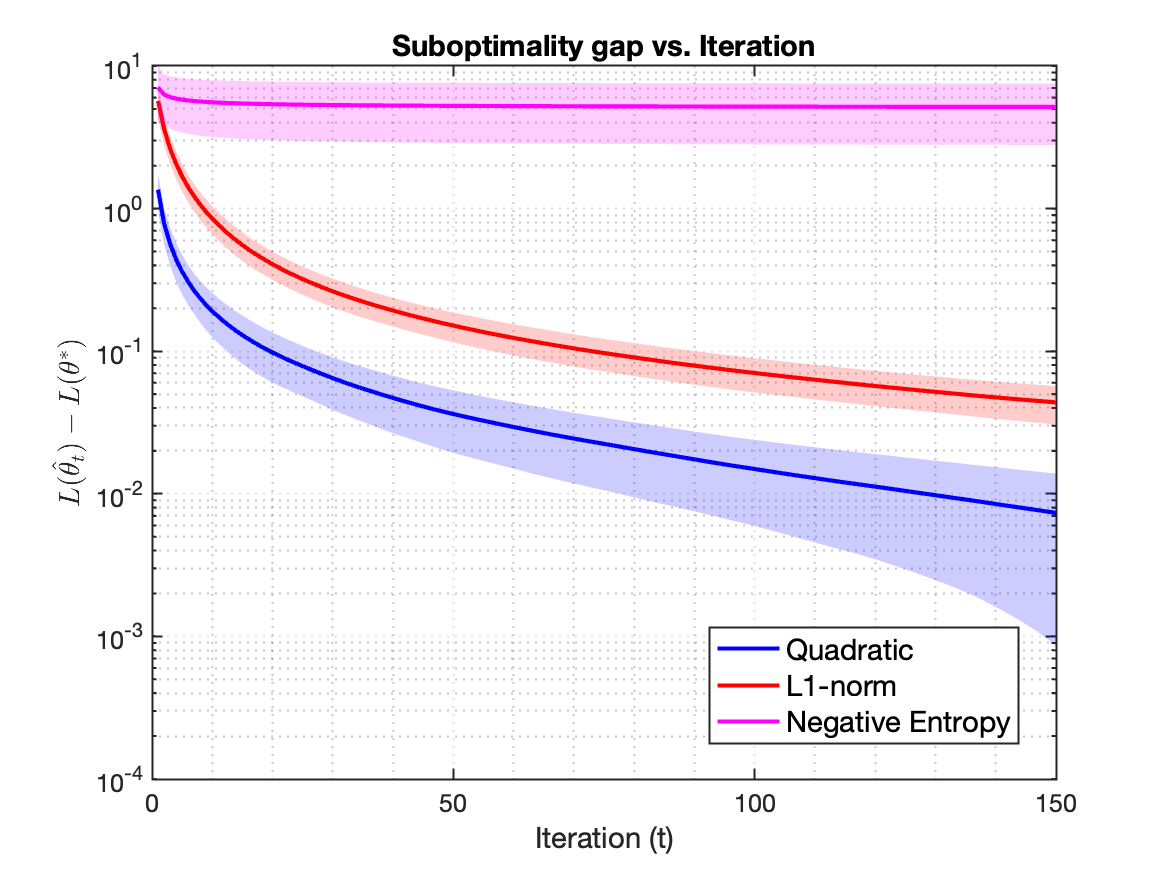}
    \caption{Performances of different mirror maps  $\psi$ for least squares regression}\label{fig:mirror_maps}
\end{figure}

Since the effectiveness of  the mirror  descent procedure relies  on the choice of the mirror map $\psi$, we conducted additional experiments to evaluate the performance of mirror descent under $\psi$'s. Specifically, we considered the negative entropy and the $\ell_1$-norm as mirror maps, in addition to the $\ell_2$-norm squared used in the preceding  experiments. The results, presented in Fig.~\ref{fig:mirror_maps}, demonstrate that the quadratic mirror map is the most suitable for   least squares  regression.

Next, we  verify the efficacy of the restart mechanism \rdgdRS \  (in Algorithm~\ref{alg:restart}) when the loss function is smooth and strongly convex. In particular, we modify the loss function of least-squares regression by adding an $\ell_2$-squared regularization term, i.e., $L(x,y;\theta)=\frac{1}{2}(x^\top\theta-y)^2+\frac{\lambda}{2}\|\theta\|_2^2$, where we set $\lambda=0.01$. Now, we run \rdgdRS \ for this strongly convex loss function.  We set  the strongly convex parameter  as $\alpha=\lambda_{\mathrm{min}}(\frac{1}{N}X^\top X+\lambda \bI_p)$ and the smooth parameter  as $M=\lambda_{\mathrm{max}}(\frac{1}{N}X^\top  X+\lambda \bI_p)$  respectively. The variance of the Gaussian noises in the uplink and downlink is set to $\sigma^2=0.1$. The function  $\psi(x)=\frac{\alpha}{2}\|x\|_2^2$.

\begin{figure}
    \centering
    \includegraphics[width=9cm]{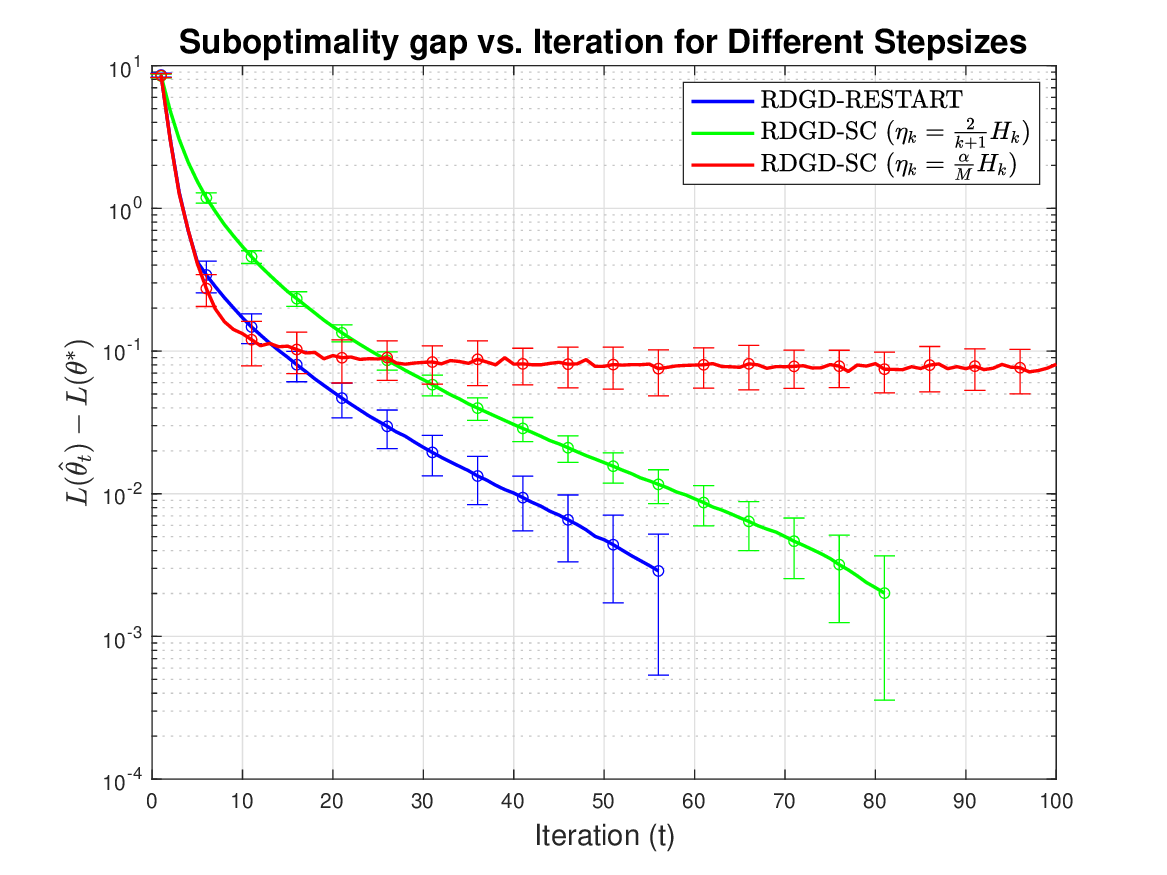}
    \caption{Comparison of  the performances of \rdgdSC \ with different stepsizes $\{\eta_k\}$ to \rdgdRS  }\label{fig:linear_strong}
\end{figure}

As shown in Fig.~\ref{fig:linear_strong}, if one chooses the stepsize sequence  $\eta_k=\frac{\alpha}{M}H_k$, the adversarial corruption accumulates after a certain point, resulting in a stagnation of the suboptimality gap.    However, it is also worth noting that the gap decreases much more rapidly when $\eta_k = \frac{\alpha}{  M}H_k$ compared to when $\eta_k = \frac{2 }{k+1}H_k$  initially. In our proposed algorithm with a restart mechanism \rdgdRS, the   gap decreases rapidly until a transition time $t_0$ which can be {\em   determined analytically}. After time $t_0+1$, we switch the stepsize sequence from $\eta_k = \frac{\alpha}{  M}H_k$ to $\eta_k = \frac{2 }{k+1}H_k$. As can be seen, the gap subsequently decreases with $t$. Our proposed algorithm  \rdgdRS \ converges faster than the $\eta_k = \frac{2 }{k+1}H_k$  case throughout the entire process, indicating a smaller time complexity to achieve the same  suboptimality gap. The experimental results presented in Fig.~\ref{fig:linear_strong} thus corroborate our theoretical analysis in Section~\ref{sec:cor_reduction}. Hence, our proposed algorithm with a restart mechanism \rdgdRS \  successfully takes advantage of the exponential decrease in the gap until corruption accumulation occurs,  then mitigates the adverse effects of the accumulation by switching to another stepsize sequence.

\subsection{Classification using L2 Support Vector Machines}
Next we verify the efficacy of \rdgd \  on a binary classification task with a loss function that is   different from that used in the previous section. Specifically, we consider  L2-SVM\footnote{We consider L2-SVM instead of the usual L1-SVM since the loss function is smooth for the former.}~\cite{chang2008coordinate, tang2013} and we use the accuracy of the binary classification task as our performance metric. The label \( y_i \) is generated such that it takes values from the set \(\{1, -1\}\) with equal probability. For each \( y_i \), the corresponding feature vector \(x_i\) is generated from a multivariate normal distribution. If \( y_i = 1 \), \(x_i\) is drawn from \(\mathcal{N}(\bm{\mu}, \gamma   \bI_p)\), and if \( y_i = -1 \), \(x_i\) is drawn from \(\mathcal{N}(-\bm{\mu}, \gamma  \bI_p)\), where \(\bm{\mu}\) is the vector of  all ones of length \( p \) and \(\gamma =4 \). % is set to~$4$.

We generated    $N=10,000$ data samples which are then split into $8,000$ samples for training and $2,000$ samples for testing. The primal objective function used in L2-SVM is  $L(x,y;\theta)  =\max \{0,1-y(x^\top \theta)\}^2 + \frac{\lambda}{2}\|\theta\|_2^2$, where $\lambda$ is set to $0.1$. We compare \rdgd \ with vanilla \dgd \ under varying levels of corruption budgets \(C(T)\), with a fixed noise variance \(\sigma^2 = 1\) in the (uplink and downlink) communication channels. Specifically, the total corruption bound up to time \(T\) is set as \(C(T) = 20  T^r\), where \(r \in \{0.25, 0.3, 0.35\}\). The total corruption budget is uniformly allocated to each iteration and worker, i.e., $c(t)=C(T)/\sqrt{T}$, for all $t\in [ T]$.

\begin{figure}
    \centering
    \includegraphics[width=9cm]{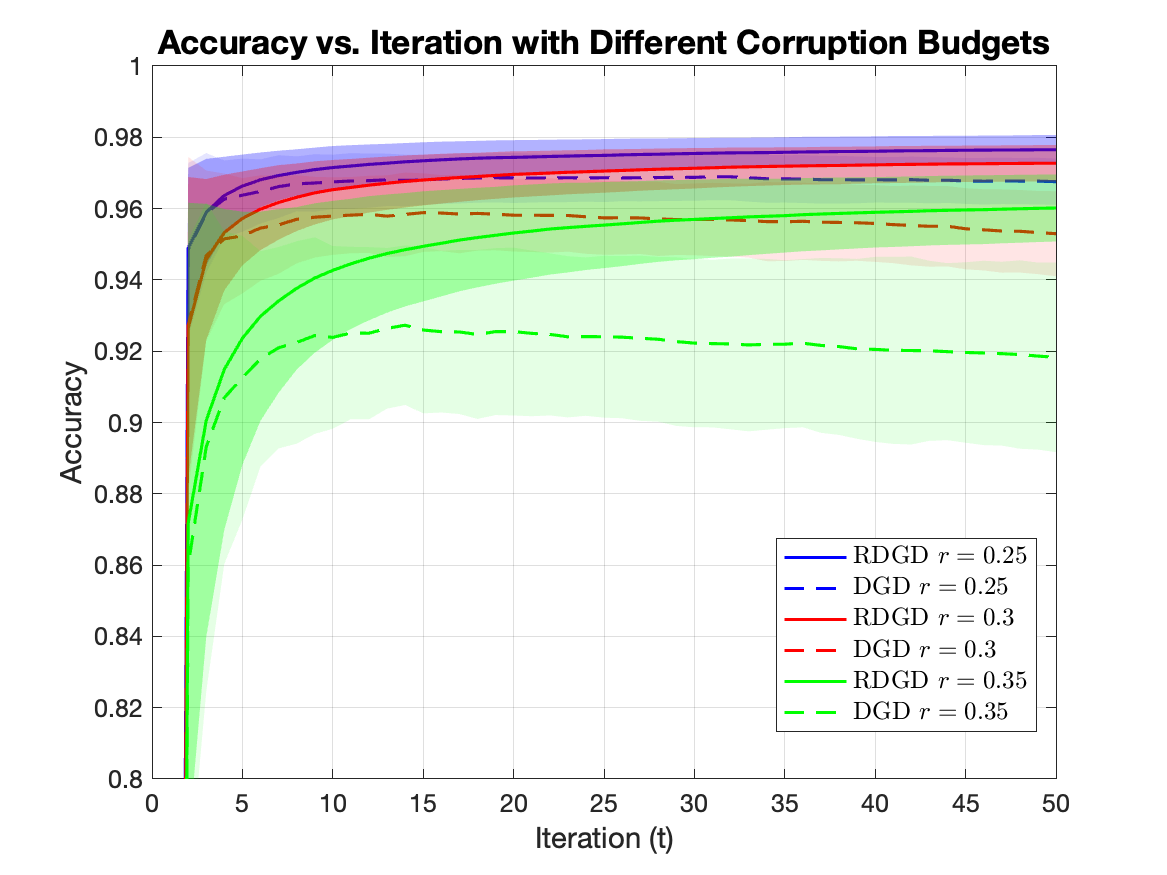}
    \caption{\rdgd\ for L2-SVM classification}\label{fig:svm_smooth}
\end{figure}

\begin{figure}
    \centering
    \includegraphics[width=9cm]{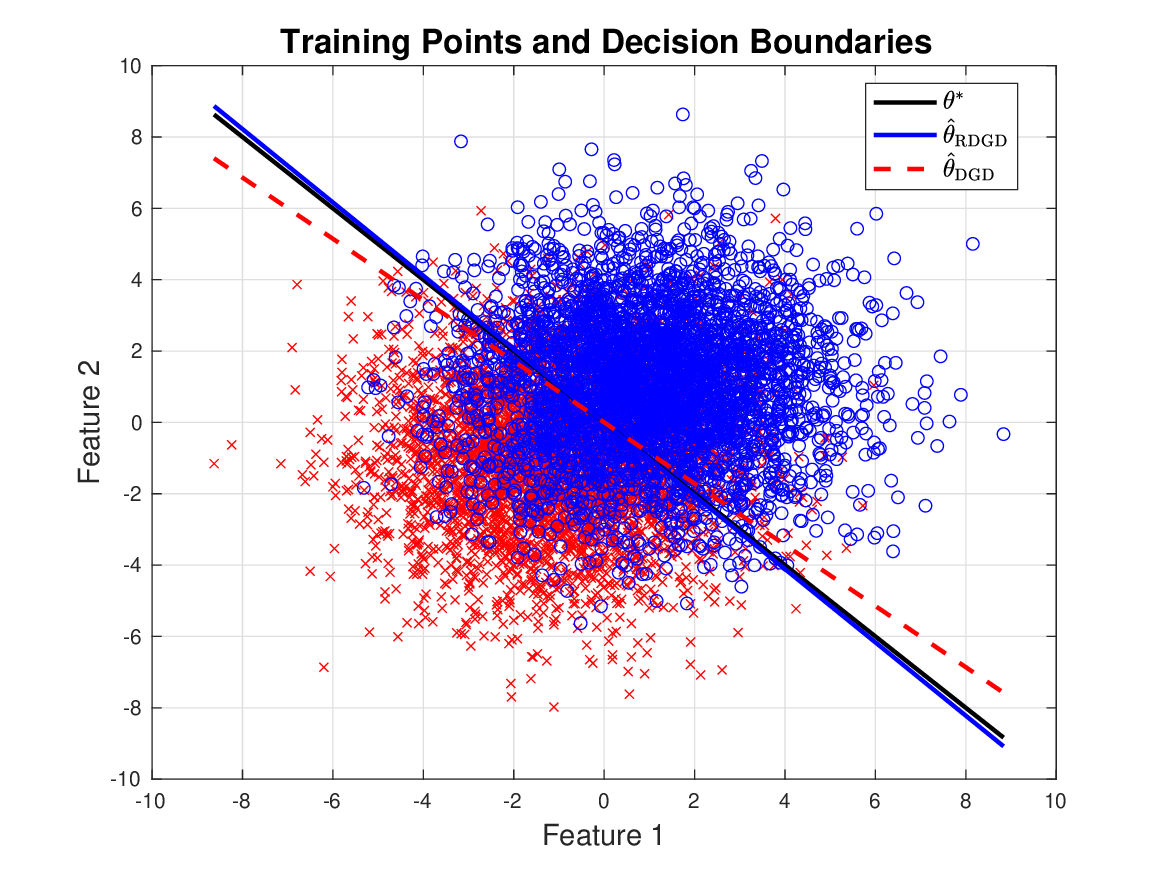}
    \caption{Decision Boundaries for L2-SVM classification }\label{fig:svm_points}
\end{figure}

As shown in Fig.~\ref{fig:svm_smooth},  \rdgd\ converges   rapidly and achieves high accuracy at any level of the corruption budget. However, the accuracy of the classification of the vanilla DGD algorithm is lower than \rdgd\ and is highly sensitive to the level of the corruption budget. \dgd\ fails when the corruption budget exceeds a certain level (e.g., $r=0.35$). In Fig.~\ref{fig:svm_points}, we display the decision boundaries for the first two dimensions of the parameter vectors $\hat{\theta}_{\rm{RDGD}}$  and $\hat{\theta}_{\rm{DGD}}$  obtained by  \rdgd\ and \dgd\ respectively when $r=0.35$. Note that $\hat{\theta}_{\rm{RDGD}}$ is almost identical to $\theta^*$. However,  $\hat{\theta}_{\rm{DGD}}$  deviates significantly from $\theta^*$, showing that \dgd\ is not robust.

\subsection{Softmax Classification on the MNIST Dataset}
In our final set of experiments, we consider a multi-class classification task on the  classical MNIST dataset, which contains 10 classes (i.e., the number of values that each label $y_i$ can take on is 10), 60,000 training samples and 10,000 test samples. The performance metric is the classification accuracy on the test samples. The dimensionality $p$ of the data points is $784=28\times28$  and the  number of workers \( m \) is set to $20$. We use the {\em cross-entropy loss} to optimize the {\em softmax classifier}.
\begin{figure}
    \centering
    \includegraphics[width=9cm]{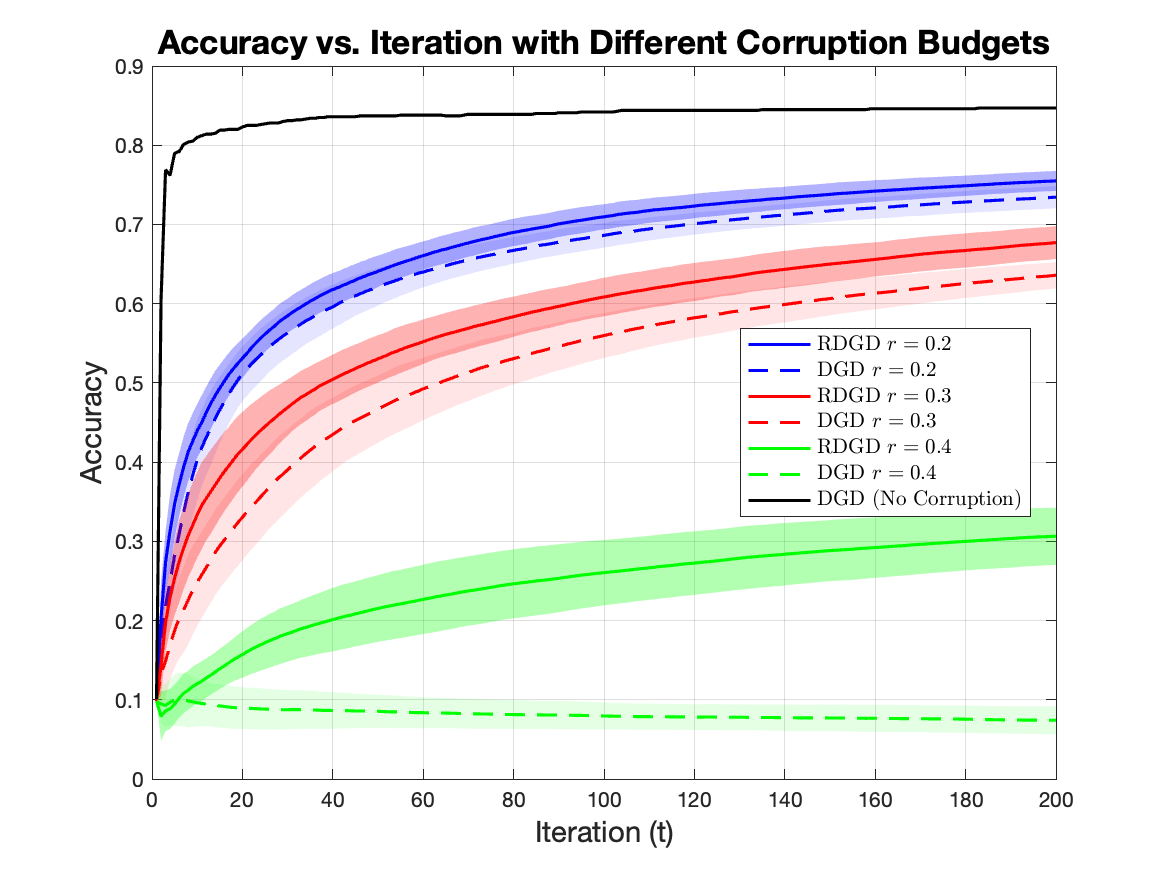}
    \caption{\rdgd\ for the MNIST dataset with different levels of corruption}\label{fig:mnist_corruption}
\end{figure}

\subsubsection{Performance under a bounded amount of adversarial corruptions}
In Fig.~\ref{fig:mnist_corruption}, we compare \rdgd \ with vanilla \dgd \ under varying levels of corruption budgets \(C(T)\), with a fixed noise variance \(\sigma^2 = 0.5\) in the  channels. Specifically, the total corruption bound up to time \(T\) is set as \(C(T) = 100  T^r\), where \(r \in \{0.2, 0.3, 0.4\}\). The total corruption budget is also uniformly allocated to each iteration and each worker as in the previous subsection. As shown in Fig.~\ref{fig:mnist_corruption}, \rdgd\  outperforms the traditional \dgd \ algorithm across all levels of the corruption budget. Furthermore, we observe that \rdgd\ achieves higher accuracy as the value of \(r\) and hence,  the amount of corruption, decreases. The vanilla   \dgd \ 
algorithm reduces to random guessing (success rate $\approx 10\%$) when the corruption budget exceeds a certain level (i.e., $r=0.4$). 

\begin{figure}
    \centering
    \includegraphics[width=9cm]{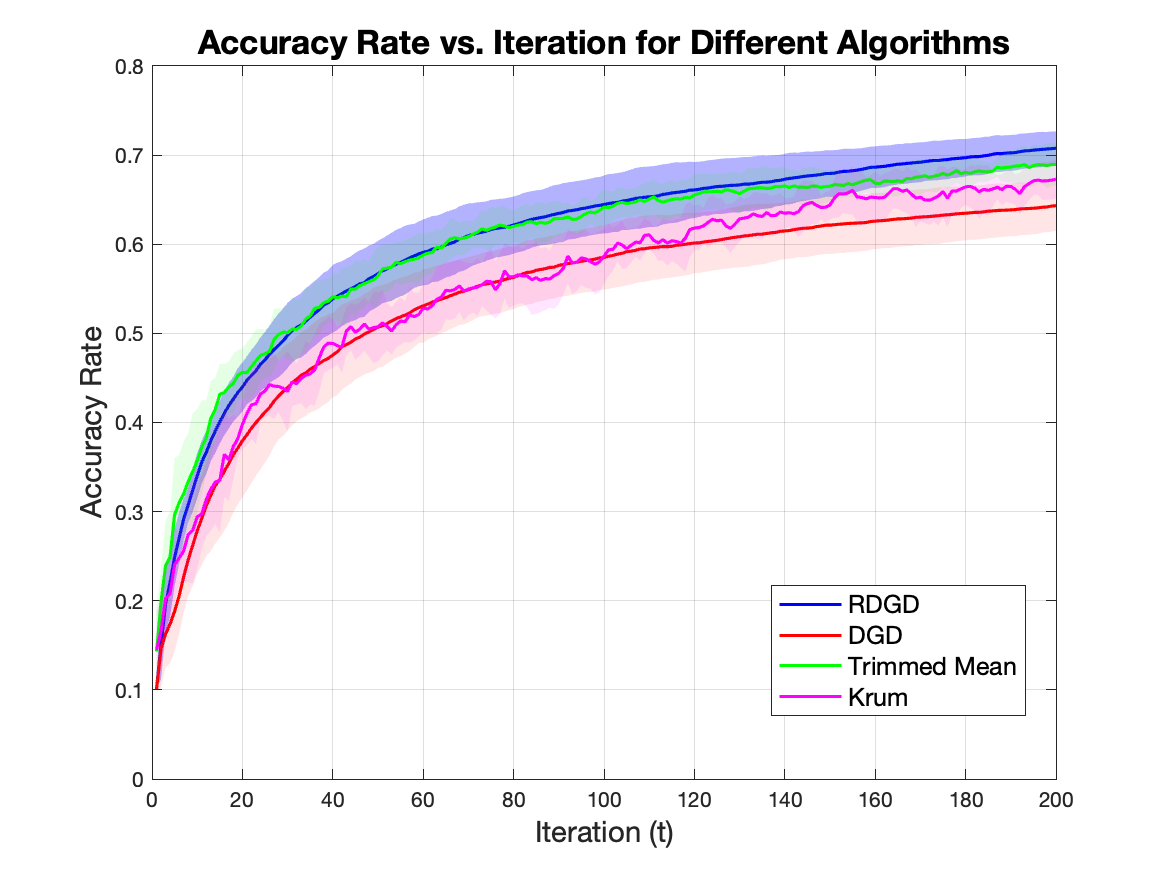}
    \caption{Accuracy rates of different algorithms for the MNIST dataset with adversarial corruption}\label{fig:comp_trimmedkrum}
\end{figure}

To further evaluate the performance of  \rdgd, we conducted experiments comparing it against three baseline algorithms: the traditional \dgd \ algorithm, Trimmed Mean by Yin {\em et al.}~\cite{yin2018byzantine}, and Krum by Blanchard {\em et al.}~\cite{blanchard2017machine}. The total corruption bound up to time $T$ and the fraction of Byzantine workers are set to be $C(T)= 150T^{0.3}$ and  $0.3$ respectively. Note that the baseline algorithms (Trimmed Mean and Krum) are only applicable when the fraction of Byzantine workers is less than half, whereas \rdgd\ can handle an arbitrary proportion.  The total
corruption budget is  \emph{uniformly allocated} to each iteration
and each Byzantine worker. The results, shown in Fig.~\ref{fig:comp_trimmedkrum}, indicate that  RDGD  consistently outperformed the baseline methods over all iterations. RDGD achieved a faster convergence rate and accuracy compared to DGD, Trimmed Mean, and Krum. 

\subsubsection{Performance under adversarial corruption in the final iterations}
\begin{figure}
    \centering
    \includegraphics[width=9cm]{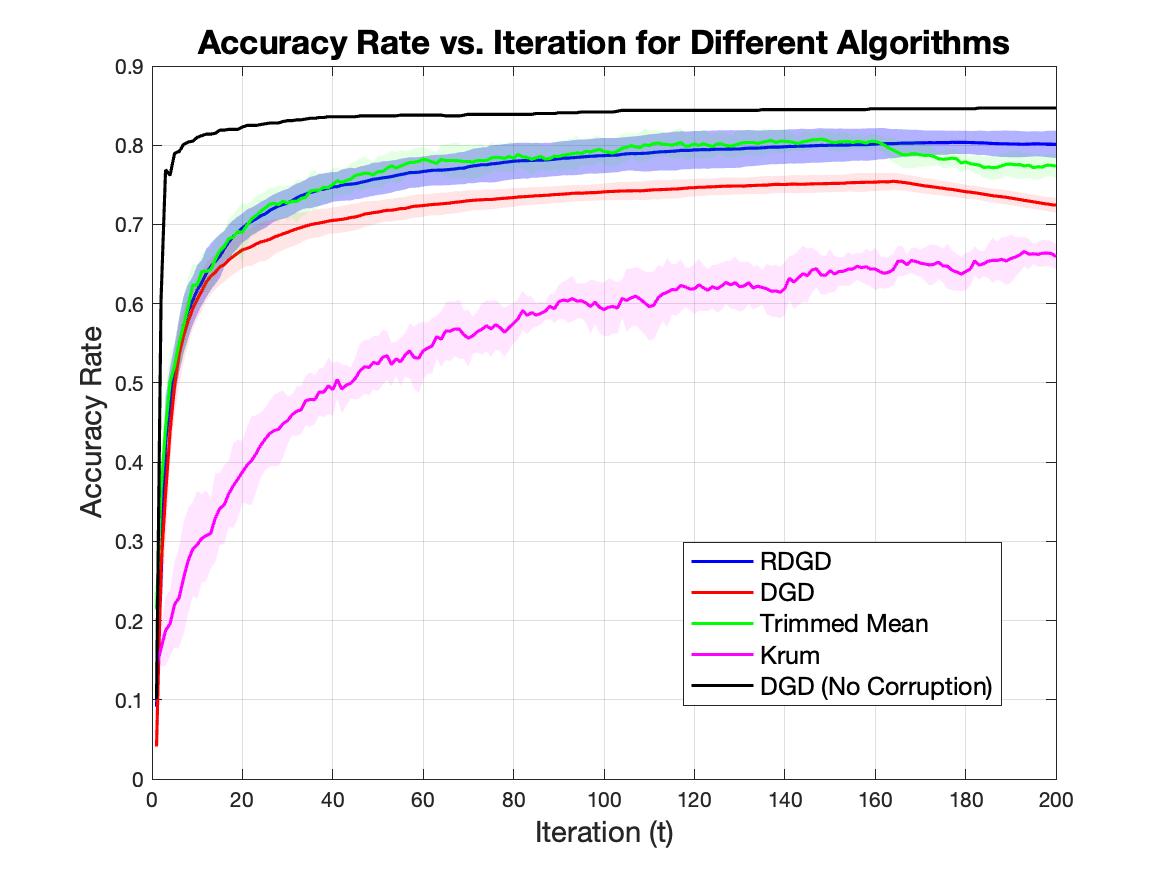}
    \caption{Accuracy rates for different algorithms for the MNIST dataset with allocating corruptions to last $20\%$ of iterations}\label{fig:last20}
\end{figure}

As mentioned, it is difficult to find the worst case
corruption pattern. In addition to allocating the available corruption budget to each iteration and each Byzantine worker uniformly, we also evaluate the performance of \rdgd\ and the baselines when the corruption is only allocated to {\em the last $20\%$ of iterations}; this is particularly challenging for all algorithms as they have to suddenly adapt to the presence of the corruptions towards the end of the horizon. The total corruption bound up to time $T$  and the fraction of Byzantine workers are set to  $C(T)= 150T^{0.3}$ and  $0.3$ respectively. Fig.~\ref{fig:last20} illustrates the performance of different algorithms when the corruption is introduced only during the last 20\% of the iterations (i.e., from  $161$ to $200$). 
In the first $160$ iterations, when no corruption is present, all algorithms exhibit   steady increases in accuracy, with \rdgd\ demonstrating the fastest convergence and the highest accuracy. However, in the last 20\% of the iterations, the malicious gradients degrade the performance of the baselines, causing noticeable drops in their accuracy, whereas \rdgd\ maintains its robustness. 

\subsubsection{Performance under periodic adversarial corruption}

\begin{figure}[h]
        \centering
        \includegraphics[width=9cm]{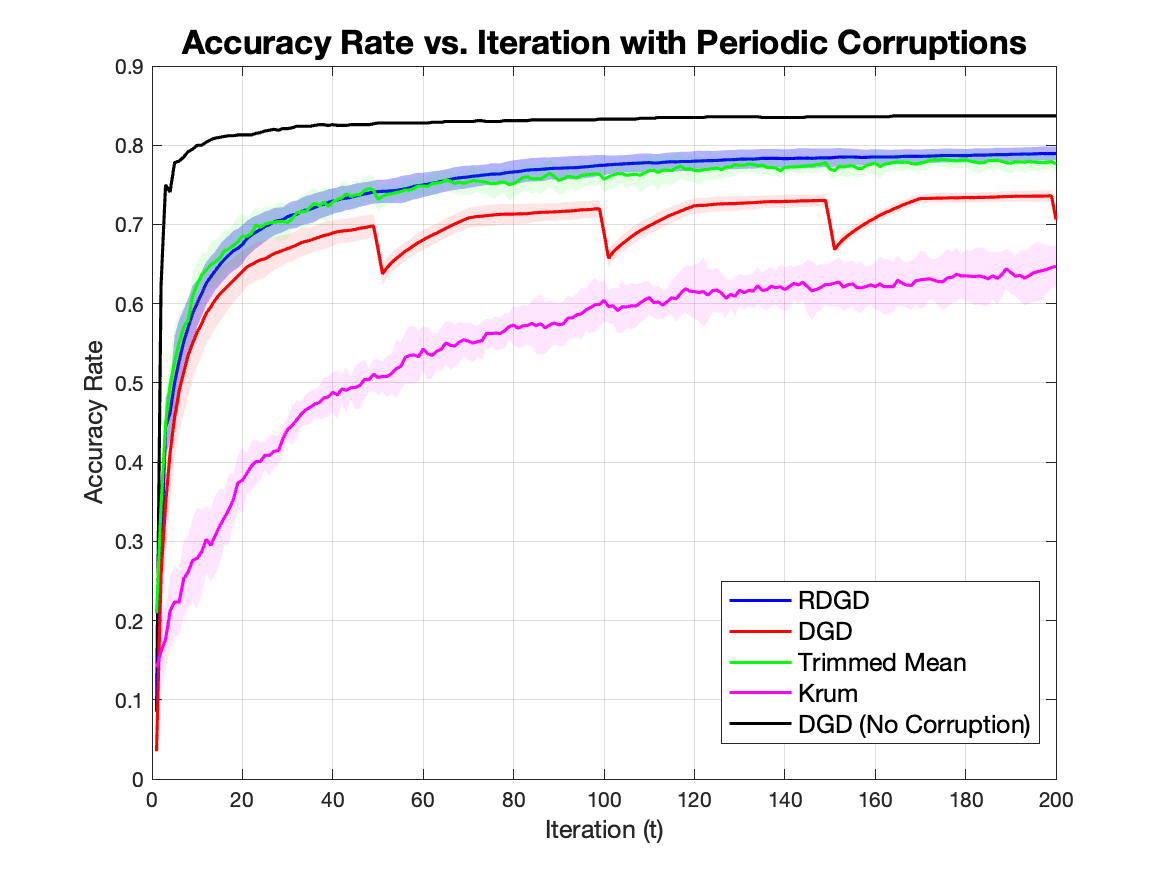}
        \caption{Accuracy rates for different algorithms for the MNIST dataset with periodic corruption}\label{fig:period}
        \end{figure}

To further show it is difficult to find the worst-case corruption pattern, we consider periodic adversarial corruptions, where the corruption is introduced at the $[50,100,150, 200]$-th iterations. The total corruption bound up to time $T$  and the fraction of Byzantine workers are set to  $C(T)= 150T^{0.3}$ and  $0.3$ respectively. Hence, the corruption level at these $4$ iterations is large given that we keep the corruption level the same at $C(T)=150T^{0.3}$. The sign of adversarial corruption at each Byzantine worker $\sign(\varepsilon_{i,t})$ is set to be $-\sign(\tilde{g}_{i,t})$. Then we add the corruption vector $\varepsilon_{i,t}$ to $g'_{i,t}$, and obtain $\bar{g}_{i,t}=g'_{i,t}+\varepsilon_{i,t}$. We checked the signs of each of the dimensions of $\bar{g}_t\in\mathbb{R}^p$ and notice that   for each Byzantine worker, more than $90\%$ of the dimensions of $\bar{g}_t$ have their signs flipped. This shows that the corruption added significantly alters the direction of the gradient vector $\bar{g}_t$, implying that its effect is substantial.
As shown in Fig.~\ref{fig:period}, \dgd\ exhibits a distinctive ``sawtooth'' pattern, sharp periodic drops in accuracy, followed by gradual recovery. These periodic drops occur when corruptions are introduced, demonstrating  \dgd's   vulnerability to ``sudden'' Byzantine attacks. Meanwhile, the performance of  \rdgd\ remains high throughout the whole process. This experiment again demonstrates  \rdgd's robustness. 

\subsubsection{Performance under ``A Little Is Enough'' attack}
Besides  adversarial corruption with a fixed budget, we also evaluate \rdgd\ for the ``A Little Is Enough'' (ALIE) attack proposed by Baruch {\em et al.}~\cite{baruch2019little}, which is designed to destroy robust aggregation methods such as Trimmed Mean and Krum. We denote the set of Byzantine workers as $\cB$. ALIE attacks the gradients at the Byzantine worker $i\in\cB $ by setting their updates to be $ g_{i,t} = \mu_{t} + z \cdot \sigma_{t} $, where $ \mu_{t} $ and \( \sigma_{t} \) are respectively the mean and standard deviation of the gradients at iteration $t$ among all the $m$ workers, and $z$ is an attack intensity parameter. This allows adversarial updates to remain statistically plausible, making them extremely difficult to detect. We set $z=1.5$ and the fraction of Byzantine workers to be $0.3$.

\begin{figure}
    \centering
    \includegraphics[width=9cm]{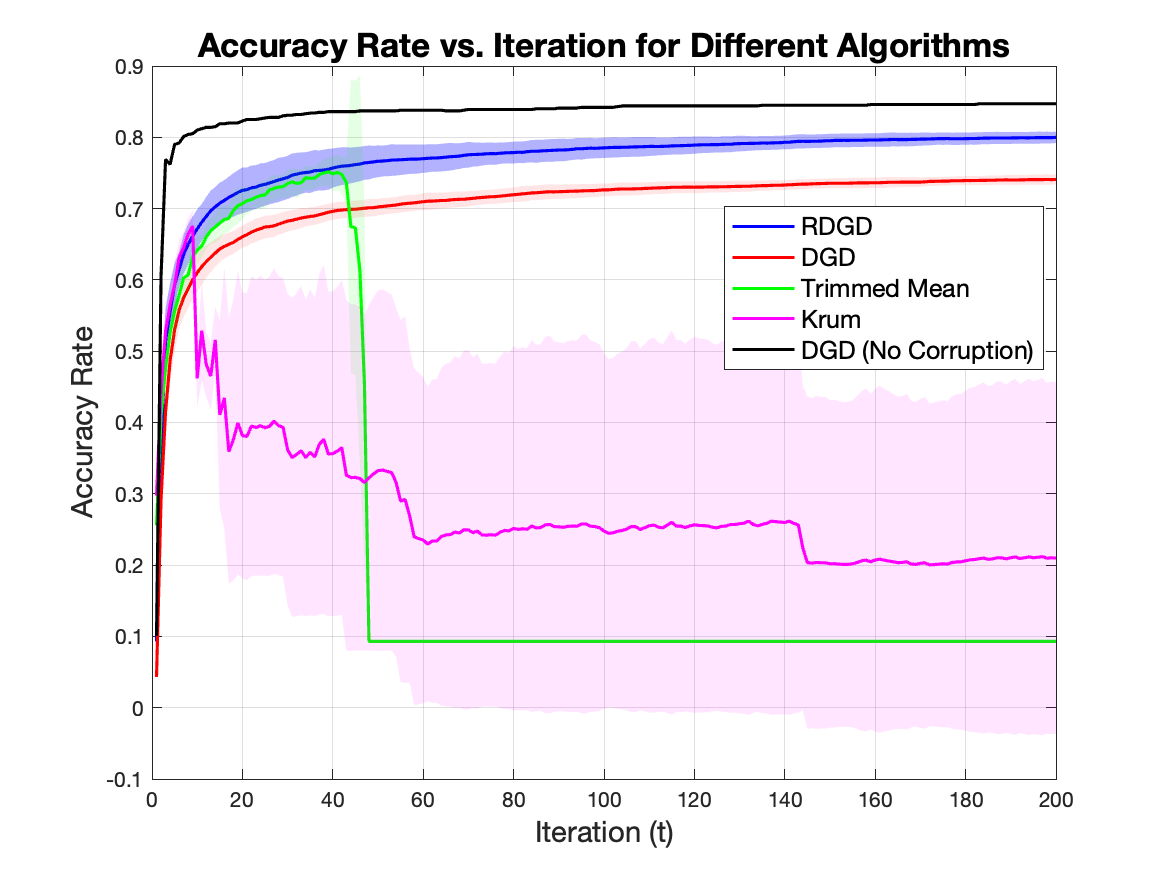}
    \caption{Accuracy rates for different algorithms for the MNIST dataset with ALIE attack}\label{fig:little}
\end{figure}

As shown in Fig.~\ref{fig:little}, \rdgd\ exhibited remarkable robustness against this attack, achieving a consistently accuracies. In contrast, baseline methods such as Trimmed Mean and Krum showed significant drops in accuracy after the early iterations.

\iffalse\begin{figure}
    \centering
    \includegraphics[width=9cm]{figs/noise1_mnist.eps}
    \caption{\rdgd\ for MNIST dataset with different levels of noise variances}\label{fig:mnist_noise}
\end{figure}\fi
\iffalse
\begin{figure}
    \centering
    \includegraphics[width=9cm]{figs/noise2_mnist.eps}
    \caption{\rdgd\ for the MNIST dataset with different   noise variances}\label{fig:mnist_noise}
\end{figure}

Next, we fix the corruption budget to be $C(t)=20t^{0.1}$ and compare \rdgd\ to  \dgd \ under different levels of noise variances in the communication channels, namely, $\sigma^2\in\{0.1,0.5,1.0\}$. As shown in Fig.~\ref{fig:mnist_noise}, the vanilla   \dgd \ 
algorithm is unable to perform effectively at any level of noise variance since  \dgd \  cannot tolerate adversarial corruption and the accuracy drops as $t$ grows.  From Fig.~\ref{fig:mnist_noise}, we observe that our proposed \rdgd\ achieves increasing accuracy as either the iteration number increases or  the  noise variance decreases. This shows that the mirror descent mechanism in \rdgd \ enables it to be  robust to adversarial corruptions. 
\fi

\section{Conclusion}\label{sec:conclusion}
In this paper, we considered a robustification of the standard distributed gradient descent algorithm used extensively in the parallelization of massive datasets in the modern era. In addition to standard channel noises on the uplink and downlink of the communication channels, we also considered mitigating the effects of {\em adversarial corruptions} at the workers via a mirror descent-type algorithm \rdgd \ and its strongly convex extension \rdgdSC. Different convergence rates of the suboptimality gap were obtained by setting different stepsizes in  \rdgdSC. We also provided a theoretical analysis of a unified algorithm \rdgdRS \ that attains the ``best-of-both-worlds'' with regard to the choice of  the different stepsizes in  \rdgdSC. Extensive numerical experiments show that (i) \rdgd \  performs much better, in terms of the suboptimality gap   and classification accuracy, than standard \dgd\ in reducing the effect of the corruption and (ii) the restarting mechanism in \rdgdRS \  is effective in speeding up the overall convergence rate for strongly convex loss functions. 

We may consider the following avenues for future research. First,  to model the scenario in which {\em some}, but not all,  workers are   compromised, we may redefine $c_t$ in~\eqref{eqn:defct} as $c_t = \sum_{i=1}^m\|\varepsilon_{i,t} \|_2$. This encourages sparsity among the $\varepsilon_{i,t}$'s  for fixed $t$. Second, we may derive information-theoretic {\em lower bounds} on the suboptimality gap as a function of the corruption $C(T)$ to determine the tightness of the upper bounds.

% \section*{Acknowlegements}
% This research/project is supported a Singapore Ministry of Education Academic Research Fund  (AcRF) Tier 2 under grant number A-8000423-00-00 and the Singapore Ministry of Education AcRF Tier 1 under grant number A-8000189-01-00.

\begin{appendices}

\section{Proof of Lemma~\ref{lem:argmin_diff}}\label{app:argmin_diff}
\begin{IEEEproof}
    By the definition of $h_t$ in \eqref{eqn:def_ht}, we have $h_t(\theta_{t+1})-h_{t-1}(\theta_{t+1})\ge \eta_t\langle \tilde{g}_t, \theta_{t+1}-\theta_t\rangle$.

    Then, based on the definition of the Bregman divergence, we have $h_{t-1}(\theta_{t+1})-h_{t-1}(\theta_{t})=\langle \nabla h_{t-1}(\theta_{t}),\theta_{t+1}-\theta_{t}\rangle+D_{h_{t-1}}(\theta_{t+1},\theta_{t})$ and $D_{h_{t-1}}(\theta_{t+1},\theta_{t})=D_{\psi}(\theta_{t+1},\theta_{t})$. However, since $\theta_{t}=\mathrm{argmin}_{u\in\Theta}h_{t-1}(u)$, we have $\langle \nabla h_{t-1}(\theta_{t}),\theta_{t+1}-\theta_{t}\rangle\ge 0$. Thus,
    \begin{align*}
        h_t(\theta_{t+1})-h_{t-1}(\theta_{t})\ge \eta_t\langle \tilde{g}_t, \theta_{t+1}-\theta_t\rangle+D_{\psi}(\theta_{t+1},\theta_{t})
    \end{align*}
    as desired.
\end{IEEEproof}

\section{Proof of Lemma~\ref{lem:smoothchange}}\label{app:smoothchange}
\begin{IEEEproof}
    Based on Lemma~\ref{lem:argmin_diff}, we have:
    \begin{align*}
        H_tL_t-H_{t-1}L_{t-1}&\ge \eta_tL(\theta_t)-\eta_t\langle\tilde{g}_t-\nabla L(\theta_t),\theta^*-\theta_t\rangle\\&\qquad+\eta_t\langle\tilde{g}_t,\theta_{t+1}-\theta_t\rangle+D_{\psi}(\theta_{t+1},\theta_{t})\\
        &= \eta_tL(\theta_t)-\eta_t\langle\tilde{g}_t-\nabla L(\theta_t),\theta^*-\theta_{t+1}\rangle\\&\;+\eta_t\langle \nabla L(\theta_t),\theta_{t+1}-\theta_t\rangle+D_{\psi}(\theta_{t+1},\theta_{t}).
    \end{align*}
    Then, it follows that:
    \begin{multline*}
        H_tG_t-H_{t-1}G_{t-1}\le H_tL(\hat{\theta}_{t})-H_{t-1}L(\hat{\theta}_{t-1})-\eta_tL(\theta_t)\\+\eta_t\langle\tilde{g}_t-\nabla L(\theta_t),\theta^*-\theta_{t+1}\rangle-\eta_t\langle \nabla L(\theta_t),\theta_{t+1}-\theta_t\rangle\\-D_{\psi}(\theta_{t+1},\theta_{t}).
    \end{multline*}

    From \rdgd \ in Algorithm~\ref{alg:smooth_mirror} and the convexity of $L$, we can write $H_tL(\hat{\theta}_{t})-H_{t-1}L(\hat{\theta}_{t-1})-\eta_tL(\theta_t)$ as:
    \begin{align}\label{zero_smooth}
        &\eta_tL(\hat{\theta}_{t})+H_{t-1}(L(\hat{\theta}_{t})-L(\hat{\theta}_{t-1}))-\eta_tL(\theta_t)\nonumber\\
        &\le \eta_t(L(\hat{\theta}_{t})-L(\theta_t))+H_{t-1}\langle\nabla L(\hat{\theta}_{t}),\hat{\theta}_{t}-\hat{\theta}_{t-1}\rangle\nonumber\\
        &=\eta_t(L(\hat{\theta}_{t})-L(\theta_t))-\eta_t\langle\nabla L(\hat{\theta}_{t}),\hat{\theta}_t-\theta_{t}\rangle\nonumber\\
        &\le \eta_t\langle\nabla L(\hat{\theta}_{t}),\hat{\theta}_t-\theta_{t}\rangle-\eta_t\langle\nabla L(\hat{\theta}_{t}),\hat{\theta}_t-\theta_{t}\rangle\nonumber\\
        &=0.
    \end{align}
    Hence, we conclude that
    \begin{multline*}
        H_tG_t-H_{t-1}G_{t-1}\le \eta_t\langle\tilde{g}_t-\nabla L(\theta_t),\theta^*-\theta_{t+1}\rangle\\-\eta_t\langle \nabla L(\theta_t),\theta_{t+1}-\theta_t\rangle-D_{\psi}(\theta_{t+1},\theta_{t})
    \end{multline*}
    as desired.
\end{IEEEproof}

\section{Proof of Theorem~\ref{thm:smooth}}\label{app:smoothmain}
\begin{IEEEproof}
    Based on Lemma~\ref{lem:smoothchange}, we have
    \begin{multline*}
    H_tG_t-H_{t-1}G_{t-1}\le \eta_t\langle\tilde{g}_t-\nabla L(\theta_t),\theta^*-\theta_{t+1}\rangle\\-\eta_t\langle \nabla L(\theta_t),\theta_{t+1}-\theta_t\rangle-D_{\psi}(\theta_{t+1},\theta_{t}).
    \end{multline*}
    By the definition of the Bregman divergence and the $\mu$-strong convexity of the mirror map $\psi$, we have $D_{\psi}(\theta_{t+1},\theta_{t})\ge \frac{\mu}{2}\|\theta_{t+1}-\theta_{t}\|_2^2$. Hence,
    \begin{multline*}
        -\eta_t\langle \nabla L(\theta_t),\theta_{t+1}-\theta_t\rangle-D_{\psi}(\theta_{t+1},\theta_{t})\\\le \eta_t K\|\theta_{t+1}-\theta_{t}\|_2-\frac{\mu}{2}\|\theta_{t+1}-\theta_{t}\|_2^2
        \le \frac{\eta_t^2K^2}{2\mu},
    \end{multline*}
    where the first inequality follows from the Cauchy--Schwarz inequality and the fact that a differentiable $K$-Lipschitz continuous and convex function $L(\cdot)$ has a gradient vector $\nabla L(\cdot)$ with norm bounded by $K$~\cite[Lemma 2.6]{shalev2012online}, and the second inequality follows from the fact that $2ab-b^2\le a^2$.

    Recall that  $g_{i,k}=\frac{1}{|\cZ_i|}\sum_{(x,y) \in \cZ_i} \nabla L(x, y; \theta_{k})$. Hence, we can write $\nabla L(\theta_{k})=\frac{1}{m}\sum_{i=1}^mg_{i,k}$ and % $\langle \tilde{g}_{k}-\nabla L(\theta_k), \theta^* - \theta_{k+1}\rangle$ as

\begin{align*}
    &\langle \tilde{g}_{k}-\nabla L(\theta_k), \theta^* - \theta_{k+1}\rangle\\&
    \!=\!\bigg\langle \frac{1}{m}\Big(\!\sum_{i=1}^mg'_{i,k}\!+\!\sum_{i=1}^m\varepsilon_{i,k}\!+\!\sum_{i=1}^mw_{k}^{(i)}\Big) -\frac{1}{m}\!\sum_{i=1}^mg_{i,k}, \theta^*\!-\! \theta_{k+1}\bigg\rangle\\
    &=\frac{1}{m}\bigg\langle\!\sum_{i=1}^m\!g'_{i,k}\!-\!\sum_{i=1}^m\!g_{i,k},\theta^*\! -\! \theta_{k+1}\!\bigg\rangle\!+\!\frac{1}{m}\bigg\langle\!\sum_{i=1}^m\!\varepsilon_{i,k},\theta^*\!-\! \theta_{k+1}\!\bigg\rangle\nonumber\\*
    & \hspace{2cm}+\frac{1}{m}\bigg\langle\sum_{i=1}^mw_k^{(i)},\theta^* - \theta_{k+1}\bigg\rangle.
\end{align*}
Recall that $R_{\Theta}=\max_{\theta\in\Theta}\|\theta-\theta^*\|_2$. It follows that
\begin{align}
&\langle \tilde{g}_{k}-\nabla L(\theta_k), \theta^* - \theta_{k+1}\rangle \nonumber\\*
&\le \frac{MR_{\Theta}}{m}\sum_{i=1}^m\|v_{k}^{(i)}\|_2+\frac{c_kR_{\Theta}}{m}+\frac{R_{\Theta}}{m}\sum_{i=1}^m\|w_k^{(i)}\|_2.
\end{align}
Then, we have
\begin{multline}\label{eq:corruption_bound}
\sum_{k=1}^t \eta_k\langle \tilde{g}_{k}-\nabla L(\theta_k), \theta^* - \theta_{k+1}\rangle\\\leq \frac{R_{\Theta}}{m}\sum_{k=1}^t \eta_{k}c_k+\frac{MR_{\Theta}}{m}\sum_{k=1}^t \eta_{k}\sum_{i=1}^m\|v_{k}^{(i)}\|_2\\+\frac{R_{\Theta}}{m}\sum_{k=1}^t \eta_{k}\sum_{i=1}^m\|w_k^{(i)}\|_2.
\end{multline}
Taking the expectation on both sides of \eqref{eq:corruption_bound} and assuming $\sigma^2_k\le \sigma^2$ for all  $k\ge 1$, we obtain
\begin{align}
     &\mathbb{E}\bigg[\sum_{k=1}^t \eta_k\langle \tilde{g}_{k}-\nabla L(\theta_k), \theta^* - \theta_{k+1}\rangle\bigg] \nonumber\\*
     &\le \frac{R_{\Theta}}{m}\langle \boldsymbol{\eta}_t,\bc_t \rangle\!+\!(M+1)\sqrt{p}\sigma R_{\Theta}\sum_{k=1}^t \eta_{k},
\end{align}
where $\boldsymbol{\eta}_t=[\eta_1,\ldots,\eta_t]^\top$ and $\bc_t=[c_1,\ldots, c_t]^\top$.

Since the overall bound on the total corruption up to and including time $t$ is $C(t)$ such that $\sqrt{\sum_{k=1}^t c_k^2}\le C(t)$, by the Cauchy--Schwarz inequality, we have
\begin{align}
&\mathbb{E} \bigg[\sum_{k=1}^t \eta_k\langle \tilde{g}_{k}-\nabla L(\theta_k), \theta^* - \theta_{k+1}\rangle\bigg]  \nonumber\\*
&\le \frac{R_{\Theta}C(t)}{m}\sqrt{\sum_{k=1}^t \eta_k^2}+(M+1)\sqrt{p}\sigma R_{\Theta}H_t.\label{eq:expect_gap}
\end{align}
Next, we bound $H_1 G_1$ as follows:
    \begin{align*}
        &H_1G_1=H_1U_1-H_1L_1\\&\le -\eta_1\langle \nabla L(\theta_1),\theta_2-\theta_1\rangle -D_{\psi}(\theta_2,\theta_0)+D_{\psi}(\theta^*,\theta_0)\\&\qquad\qquad\qquad+\eta_1\langle \tilde{g}_{1}-\nabla L(\theta_1), \theta^*-\theta_2\rangle\\&\le \frac{\eta_1^2K^2}{2\mu}+D_{\psi}(\theta^*,\theta_0)+\eta_1\langle \tilde{g}_{1}-\nabla L(\theta_1), \theta^*-\theta_2\rangle.
    \end{align*}
Then, we conclude that
    \begin{multline*}
        \mathbb{E}[L(\hat{\theta}_t)-L(\theta^*)]\le \frac{D_{\psi}(\theta^*,\theta_0)}{H_t}+\frac{K^2}{2\mu}\cdot\frac{\sum_{k=1}^t \eta_k^2}{H_t}\\+\frac{R_{\Theta}C(t)}{m}\cdot\frac{\sqrt{\sum_{k=1}^t \eta_k^2}}{H_t}+(M+1)\sqrt{p}\sigma R_{\Theta}.
    \end{multline*}
Recall that $T$ is the  fixed and known number of iterations. Let  $\eta_k=\frac{1}{\sqrt{T}}$ for all $k\ge1$. Finally we obtain the desired bound as follows:
\begin{multline*}
    \mathbb{E}[L(\hat{\theta}_T)-L(\theta^*)]\le \frac{D_{\psi}(\theta^*,\theta_0)}{\sqrt{T}}+\frac{K^2}{2\mu}\cdot\frac{1}{\sqrt{T}}\\+\frac{R_{\Theta}C(T)}{m}\cdot\frac{1}{\sqrt{T}}+(M+1)\sqrt{p}\sigma R_{\Theta}.
\end{multline*}
\end{IEEEproof}

\section{Proof of Lemma~\ref{lem:initial_lower}}\label{app:initial_lower}
\begin{IEEEproof}
    The initial lower bound $L_1$ multiplied by $H_1$ can be expressed as
  
    \begin{multline*}
    H_1L_1=\eta_1L(\theta_1)+\eta_1\langle \tilde{g}_{1}, \theta_2-\theta_1\rangle+\eta_1\alpha\|\theta_2-\theta_1\|_2^2\\-\frac{\eta_1\alpha}{2}\|\theta^*-\theta_1\|_2^2-\eta_1\langle \tilde{g}_{1}-\nabla L(\theta_1), \theta^* - \theta_1\rangle.
    \end{multline*}
The corresponding upper bound $U_1$ multiplied by $H_1$ can be written as
    \begin{align*}   
    H_1U_1=\eta_1L(\hat{\theta}_1)=\eta_1L(\theta_1).
    \end{align*}
Then, we have

    \begin{align*}
        &H_1U_1-H_1L_1=H_1G_1\\
        &\le\frac{\eta_1\alpha}{2}\|\theta^*-\theta_1\|_2^2+\eta_1\langle \tilde{g}_{1}-\nabla L(\theta_1), \theta^*-\theta_2\rangle\\&\qquad\quad-\eta_1\langle \nabla L(\theta_1), \theta_2-\theta_1\rangle-\eta_1\alpha\|\theta_2-\theta_1\|_2^2\\
        &\leq \frac{\eta_1\alpha}{2}\|\theta^*-\theta_1\|_2^2+\eta_1\langle \tilde{g}_{1}-\nabla L(\theta_1), \theta^*-\theta_2\rangle\\&\quad\quad\quad+\eta_1 K\|\theta_2-\theta_1\|_2-\eta_1\alpha\|\theta_2-\theta_1\|_2^2\\
        &\leq\frac{\eta_1\alpha}{2}\|\theta^*-\theta_1\|_2^2+\eta_1\langle \tilde{g}_{1}-\nabla L(\theta_1), \theta^*-\theta_2\rangle+\frac{\eta_1 K^2}{4\alpha}
    \end{align*}
    as desired.
\end{IEEEproof}

\section{Proof of Lemma~\ref{lem:stronglower}}\label{app:stronglower}
\begin{IEEEproof}
    By the definition of $h_t$ in~\eqref{eqn:def_ht_sc}, we have $h_t(\theta_{t+1})-h_{t-1}(\theta_{t+1})\ge \eta_t\langle \tilde{g}_t, \theta_{t+1}-\theta_t\rangle+\frac{\eta_t\alpha}{2}\|\theta_{t+1}-\theta_t\|_2^2$.
Then, based on the definition of the Bregman divergence, we have
    \begin{multline*}
        h_{t-1}(\theta_{t+1})-h_{t-1}(\theta_{t})=\langle \nabla h_{t-1}(\theta_{t}),\theta_{t+1}-\theta_{t}\rangle\\+D_{h_{t-1}}(\theta_{t+1},\theta_{t}).
    \end{multline*}
    Since $\theta_{t}=\mathrm{argmin}_{u\in\Theta}h_{t-1}(u)$, we have $\langle \nabla h_{t-1}(\theta_{t}),\theta_{t+1}-\theta_{t}\rangle\ge 0$. Then:
    \begin{align*}
        &h_{t-1}(\theta_{t+1})-h_{t-1}(\theta_{t})\nonumber\\*
        &\ge D_{h_{t-1}}(\theta_{t+1},\theta_{t})\\
        &=\frac{H_{t-1}\alpha}{2}\|\theta_{t+1}-\theta_{t}\|_2^2+\frac{\eta_1\alpha}{2}\|\theta_{t+1}-\theta_{t}\|_2^2\\
        &\geq \frac{H_{t-1}\alpha}{2}\|\theta_{t+1}-\theta_{t}\|_2^2.
    \end{align*}
    Hence, we have
    \begin{align}
    h_t(\theta_{t+1})-h_{t-1}(
    \theta_{t}) &\ge \eta_t\langle \tilde{g}_t, \theta_{t+1}-\theta_t\rangle\\
    &\qquad+\frac{H_{t}\alpha}{2}\|\theta_{t+1}-\theta_t\|_2^2
    \end{align}
    as desired.
\end{IEEEproof}

\section{Proof of Lemma~\ref{lem:gap_strong}}\label{app:stronggap}
\begin{IEEEproof}
    From Lemma~\ref{lem:stronglower}, we have
    \begin{align*}   
    &H_tL_t-H_{t-1}L_{t-1}\\&\ge \eta_tL(\theta_t)+\eta_t\langle \tilde{g}_t,\theta_{t+1}-\theta_t \rangle+\frac{H_{t}\alpha}{2}\|\theta_{t+1}-\theta_{t}\|_2^2\\&\qquad\qquad\qquad-\eta_t\langle \tilde{g}_{t}-\nabla L(\theta_t), \theta^* - \theta_t\rangle\\
    &=\eta_tL(\theta_t)\!+\!\eta_t\langle \nabla L(\theta_t),\theta_{t+1}-\theta_t \rangle\!+\!\frac{H_{t}\alpha}{2}\|\theta_{t+1}-\theta_{t}\|_2^2\\&\qquad\qquad-\eta_t\langle \tilde{g}_{t}-\nabla L(\theta_t), \theta^* - \theta_{t+1}\rangle.
    \end{align*}
    Then, we have
    \begin{multline}\label{eq:gap_strong}
        H_tG_t-H_{t-1}G_{t-1}\le H_tL(\hat{\theta}_{t})-H_{t-1}L(\hat{\theta}_{t-1})\\-\eta_tL(\theta_t)-\eta_t\langle \nabla L(\theta_t),\theta_{t+1}-\theta_t \rangle-\!\frac{H_{t}\alpha}{2}\|\theta_{t+1}-\theta_{t}\|_2^2\\+\eta_t\langle \tilde{g}_{t}-\nabla L(\theta_t), \theta^* - \theta_{t+1}\rangle.
    \end{multline}

    Combining the fact that $\hat{\theta}_{t}=\frac{H_{t-1}}{H_t}\hat{\theta}_{t-1}+\frac{\eta_t}{H_t}\theta_t$ and~\eqref{zero_smooth}, we have $H_tL(\hat{\theta}_{t})-H_{t-1}L(\hat{\theta}_{t-1})-\eta_tL(\theta_t)\le 0$. Hence, we may rewrite the right-hand side of \eqref{eq:gap_strong} as follows
    \begin{multline*}
    -\eta_t\langle \nabla L(\theta_t),\theta_{t+1}-\theta_t \rangle-\!\frac{H_t\alpha}{2}\|\theta_{t+1}-\theta_{t}\|_2^2\\+\eta_t\langle \tilde{g}_{t}-\nabla L(\theta_t), \theta^* - \theta_{t+1}\rangle.
    \end{multline*}
    
    \iffalse
    Combining the fact that $\frac{\eta_{t}}{H_{t}}\le \frac{\alpha}{M}$ and the $M$-smoothness of $L$, we obtain
    \begin{align}\label{eq:part2_strong}
    &\eta_t\left(L(\theta_{t+1})\!-\!L(\theta_t)\!-\!\langle \nabla L(\theta_t),\theta_{t+1}\!-\!\theta_t \rangle\right)\!-\!\frac{H_t\alpha}{2}\|\theta_{t+1}-\theta_{t}\|_2^2\nonumber\\
    &\le \frac{\eta_t M}{2}\|\theta_{t+1}\!-\!\theta_{t}\|_2^2-\!\frac{H_t\alpha}{2}\|\theta_{t+1}-\theta_{t}\|_2^2=0.
    \end{align}
    By the convexity of $L$ and the Cauchy--Schwartz inequality,
    \begin{align}\label{eq:part3_strong}
    &\eta_t(L(\theta_{t})-L(\theta_{t+1}))-\!\frac{\eta_t\alpha}{2}\|\theta_{t+1}-\theta_{t}\|_2^2\nonumber\\&\le -\eta_t\langle\nabla L(\theta_t),\theta_{t+1}-\theta_t\rangle-\!\frac{\eta_t\alpha}{2}\|\theta_{t+1}-\theta_{t}\|_2^2\nonumber\\
    &\le \frac{\eta_t M^2}{2\alpha}.
    \end{align}
    \fi
    Then, we have
    \begin{align}\label{eq:strongpart}
    &-\eta_t\langle \nabla L(\theta_t),\theta_{t+1}-\theta_t \rangle-\!\frac{H_t\alpha}{2}\|\theta_{t+1}-\theta_{t}\|_2^2\nonumber\\
    &\leq \eta_t K\|\theta_{t+1}-\theta_t\|_2-\frac{H_t\alpha}{2}\|\theta_{t+1}-\theta_{t}\|_2^2\nonumber\\
    &\leq \frac{\eta_t^2 K^2}{2H_t\alpha} \leq \frac{\eta_t K^2}{2\alpha},
    \end{align}
    where the last inequality from the fact that $H_t\geq \eta_t$.
    
    Combining \eqref{eq:gap_strong} and \eqref{eq:strongpart}, we  conclude that
    \begin{equation*}
    H_tG_t\le H_{t-1}G_{t-1}+\eta_t\langle \tilde{g}_{t}-\nabla L(\theta_t), \theta^* - \theta_{t+1}\rangle+\frac{\eta_t K^2}{2\alpha}
    \end{equation*}
    as desired.
\end{IEEEproof}

\section{Proof of Corollary~\ref{cor:strongstep}}\label{app:strongstep}
\begin{IEEEproof}
When $\frac{\eta_t}{H_t}=\frac{\alpha}{M}$, we write $G_t=\frac{H_1}{H_t}G_1+\frac{\sum_{k=2}^t E_k}{H_t}$ as
\begin{align*}
G_t&=\frac{H_1}{H_2}\frac{H_2}{H_3}\dotsm\frac{H_{t-1}}{H_{t}}G_1+\frac{\sum_{k=2}^t E_k}{H_t}\\
 &=\Big(1-\frac{\eta_2}{H_2}\Big)\Big(1-\frac{\eta_3}{H_3}\Big)\dotsm\Big(1-\frac{\eta_t}{H_t}\Big)G_1+\frac{\sum_{k=2}^tE_k}{H_t}\\
 &=\Big(1-\frac{\alpha}{M}\Big)^{t-1}G_1+\frac{\sum_{k=2}^tE_k}{H_t} 
\end{align*}

From Lemmas~\ref{lem:initial_lower} and \ref{lem:gap_strong}, it follows that 
\begin{multline*}
G_t=\Big(1-\frac{\alpha}{M}\Big)^{t-1}\frac{\alpha}{2}\|\theta^*-\theta_0\|_2^2+\frac{ K^2\sum_{k=1}^t\eta_k}{2\alpha H_t}\\+\frac{\sum_{k=1}^t\eta_k\langle \tilde{g}_{k}-\nabla L(\theta_k), \theta^* - \theta_{k+1}\rangle}{H_t}
\end{multline*}

By \eqref{eq:expect_gap}, we obtain the following sequence of upper bounds on the expectation of the final term above:
\begin{align*}
&\mathbb{E}\bigg[\frac{\sum_{k=1}^t\eta_k\langle \tilde{g}_{k}-\nabla L(\theta_k), \theta^* - \theta_{k+1}\rangle}{H_t}\bigg] \\
&=\frac{R_{\Theta}C(t)}{m}\frac{\sqrt{\sum_{k=1}^t \eta_k^2}}{H_t}+(M+1)\sqrt{p}\sigma R_{\Theta}\\
&=\frac{R_{\Theta}C(t)}{m}\sqrt\frac{{\sum_{k=1}^t \eta_k^2}}{H_t^2}+(M+1)\sqrt{p}\sigma R_{\Theta}\\
&=\frac{R_{\Theta}C(t)\alpha}{mM}\sqrt{\frac{\sum_{k=1}^t H_k^2}{H_t^2}}+(M+1)\sqrt{p}\sigma R_{\Theta}\\
&=\frac{R_{\Theta}C(t)\alpha}{mM}\sqrt{ \Big(\frac{H_1}{H_t}\Big)^2+\ldots+\Big(\frac{H_t}{H_t}\Big)^2}+(M+1)\sqrt{p}\sigma R_{\Theta}\\
&\le\frac{R_{\Theta}C(t)\alpha}{mM}\sqrt{\frac{1-(1-\frac{\alpha}{M})^{2(t-1)}}{1-(1-\frac{\alpha}{M})^2}}+(M+1)\sqrt{p}\sigma R_{\Theta}\\
&\le\frac{R_{\Theta}C(t)}{m}+(M+1)\sqrt{p}\sigma R_{\Theta}.
\end{align*}
Therefore, we  conclude that:
\begin{multline*}
\mathbb{E}[L(\hat{\theta}_t)-L(\theta^*)]\le \Big(1-\frac{\alpha}{M}\Big)^{t-1}\frac{\alpha}{2}\|\theta^*-\theta_0\|_2^2+\frac{ K^2}{2\alpha}\\+\frac{R_{\Theta}C(t)}{m}+(M+1)\sqrt{p}\sigma R_{\Theta}.
\end{multline*}

When $\eta_k=\frac{2}{k+1}H_k$, the result of Case 2 follows by substituting this new choice of stepsize into Theorem~\ref{thm:strong}.
\end{IEEEproof}

\section{Proof of Lemma~\ref{lem:trans_time}}\label{app:transtime}
\begin{IEEEproof}
We   obtain the following inequality by substituting $C(t)=m t^r$ into the general bound for the smooth and strongly convex case in \eqref{eqn:smooth_sc}
    \begin{multline*}
        \mathbb{E}[L(\hat{\theta}_t)-L(\theta^*)]\le \Big(1-\frac{\alpha}{M}\Big)^{t-1}\frac{\alpha}{2}\|\theta^*-\theta_0\|_2^2+\frac{ K^2}{2\alpha}\\+R_{\Theta}t^r+(M+1)\sqrt{p}\sigma R_{\Theta}. 
    \end{multline*}
From the fact that $\|\theta^*-\theta_0\|_2^2\le R_{\Theta}^2$, we can simplify the  above upper bound of $\mathbb{E}[L(\hat{\theta}_t)-L(\theta^*)]$ as follows
    \begin{multline}\label{eq:upper_constantstep}
        \mathbb{E}[L(\hat{\theta}_t)-L(\theta^*)]\le \frac{\alpha R_{\Theta}^2}{2} e^{-\frac{\alpha}{M}(t-1)}+\frac{ K^2}{2\alpha}\\+R_{\Theta}t^r+(M+1)\sqrt{p}\sigma R_{\Theta}. 
    \end{multline}
We take the derivative of the upper bound in~\eqref{eq:upper_constantstep} with respect to $t$ and set the derivative to be zero as follows:
\begin{equation}\label{eq:derivate}
-\frac{\alpha^2 R_{\Theta}^2}{2M}e^{-\frac{\alpha}{M}(t-1)}+\frac{r\cdot R_{\Theta}}{t^{1-r}}=0.
\end{equation}
Then, we can rewrite \eqref{eq:derivate} as
\begin{align*}
   % t^{1-r}e^{-\frac{\alpha}{M}t}&=\frac{2Mr}{\alpha^2 R_{\Theta}\exp(\alpha/M)}\\
    %\Longleftrightarrow\qquad 
    t e^{-\frac{\alpha}{(1-r)M}t}&=\left(\frac{2Mr}{\alpha^2 R_{\Theta}\exp(\alpha/M)}\right)^{\frac{1}{1-r}}.
\end{align*}
Let $x=-\frac{\alpha }{(1-r)M}t$, we have
\begin{equation*}
     x e^x=-\frac{\alpha}{(1-r)M}\left(\frac{2Mr}{\alpha^2 R_{\Theta}\exp(\alpha/M)}\right)^{\frac{1}{1-r}}=: B(r).
\end{equation*}
Therefore, we can obtain the solution $t_0$ by introducing the Lambert $W$ function $W_{-1}$ as follows:
\begin{equation*}
    t_0=\left\lceil-\frac{(1-r)M}{\alpha}W_{-1}\left(B(r)\right)\right\rceil.
\end{equation*}
\end{IEEEproof}

\section{Proof of Theorem~\ref{thm:errorrestart}}\label{app:errorrestart}
\begin{IEEEproof}
Substituting $C(t)=m t^r$ into \eqref{eqn:smooth_sc}, we obtain
    \begin{multline*}
        \mathbb{E}[L(\hat{\theta}_t)-L(\theta^*)]\le \Big(\prod_{k=2}^{t} (1-\frac{\eta_k}{H_k})\Big)\frac{\alpha}{2}\|\theta^*-\theta_0\|_2^2+\frac{ K^2}{2\alpha}\\+R_{\Theta}t^r\cdot\frac{\sqrt{\sum_{k=1}^t \eta_k^2}}{H_t}+(M+1)\sqrt{p}\sigma R_{\Theta}. 
    \end{multline*}
We now decompose the factor $\prod_{k=2}^{t} (1-\frac{\eta_k}{H_k})$ as follows:
\begin{align*}
\prod_{k=2}^{t} (1-\frac{\eta_k}{H_k})&=\bigg(\prod_{k=2}^{t_0}(1-\frac{\alpha}{M})\bigg)\cdot \prod_{k=t_0+1}^{t}\Big(1-\frac{2}{k+1}\Big)\\
&=\Big(1-\frac{\alpha}{M}\Big)^{t_0-1}\cdot \frac{t_0(t_0+1)}{t(t+1)}.
\end{align*}
Next, we decompose  the term $\frac{1}{H_t}\sqrt{\sum_{k=1}^t \eta_k^2}$ as follows:
\begin{equation}\label{eq:stepsplit}
\frac{\sqrt{\sum_{k=1}^t \eta_k^2}}{H_t}\!=\!\sqrt{ \sum_{i=1}^{t_0}\Big(\frac{\eta_i}{H_t}\Big)^2+ \sum_{i=t_0+1}^t \Big(\frac{\eta_{i}}{H_t}\Big)^2 }.
\end{equation}

We now seek to bound  $\sum_{i=t_0+1}^t(\frac{\eta_{i}}{H_t})^2$ in \eqref{eq:stepsplit}. We first derive  the ratio   $(\frac{\eta_{i-1}}{H_t})^2/(\frac{\eta_{i}}{H_t})^2$ for all  $i\in\{t_0+2,\ldots,  t\}$. Noting that  $\frac{\eta_i}{H_i}=\frac{2}{i+1}$,
\begin{align}
    &\Big(\frac{\eta_{i-1}}{H_t}\Big)^2/\Big(\frac{\eta_{i}}{H_t}\Big)^2=\Big(\frac{\eta_{i-1}}{\eta_{i}}\Big)^2\nonumber\\
    &=\Big(\frac{i+1}{i}\cdot \frac{H_{i-1}}{H_{i}}\Big)^2=\Big(\frac{i+1}{i}\cdot \frac{H_{i}-\eta_{i}}{H_{i}}\Big)^2\nonumber\\
    &=\Big(\frac{i+1}{i}\cdot \left(1-\frac{\eta_{i}}{H_{i}}\right)\Big)^2=\Big(\frac{i+1}{i}\cdot \Big(1-\frac{2}{i+1}\Big)\Big)^2\nonumber\\
    &=\Big(\frac{i-1}{i}\Big)^2.\nonumber
\end{align}
Hence, we have
\begin{align}
    &\sum_{i=t_0+1}^{t }\Big(\frac{\eta_{i}}{H_t}\Big)^2\nonumber\\
    &=\Big(\frac{\eta_{t}}{H_t}\Big)^2\bigg(1\!+\!\Big(\frac{t-1}{t}\Big)^2\!+\!\Big(\frac{t-2}{t}\Big)^2\!+\!\dotsm\!+\!\Big(\frac{t_0+1}{t}\Big)^2\bigg)\nonumber\\
    &=\Big(\frac{\eta_{t}}{H_t}\Big)^2\bigg(\frac{t^2+(t-1)^2+(t-2)^2+\dotsm+(t_0+1)^2}{t^2}\bigg)\nonumber\\
    &=\Big(\frac{2}{t+1}\Big)^2\bigg(\frac{\sum_{i=t_0+1}^t i^2}{t^2}\bigg)\label{eq:sum_part2}. 
\end{align}

Next, we can bound  the remaining part in \eqref{eq:stepsplit} as follows
\begin{align}
\sum_{i=1}^{t_0} \Big(\frac{\eta_{i}}{H_t}\Big)^2 &\stackrel{(a)}=\frac{(\frac{\eta_{t_0}}{H_t})^2(1-(1-\alpha/M)^{2t_0})}{1-(1-\alpha/M)^2}\nonumber\\
&\leq \frac{(\frac{\eta_{t_0}}{H_t})^2}{1-(1-\alpha/M)^2}\stackrel{(b)}\leq \frac{M^2}{\alpha^2}\Big(\frac{\eta_{t_0}}{H_t} \Big)^2,\nonumber\label{eq:sum_part1}
\end{align}
where $(a)$ is due to the fact that $(\frac{\eta_{j}}{H_t})^2/(\frac{\eta_{j+1}}{H_t})^2=(1-\alpha/M)^2$ for all $j\in\{1, \ldots, t_0-1\}$ and $(b)$  is due to the  fact that $1-(1-\alpha/M)^2\geq \alpha^2/M^2$ when $\alpha/M\leq 1$.

To compute  $(\frac{\eta_{t_0}}{H_t})^2$,  consider 
\begin{align*}
    \frac{\eta_{t_0+1}}{H_{t_0+1}} &=\frac{2}{t_0+2}  , &\quad
    \frac{\eta_{t_0+1}}{H_{t_0}+\eta_{t_0+1}}&=\frac{2}{t_0+2}, \\
    \frac{H_{t_0}}{\eta_{t_0+1}}&=\frac{t_0}{2}, &\quad
     \frac{\eta_{t_0}}{\eta_{t_0+1}} &=\frac{\alpha}{M}\cdot\frac{t_0}{2},
\end{align*}
where the final equality is because $\eta_{t_0}/H_{t_0}=\alpha/M$.

From $\big(\frac{\eta_{t_0+1}}{H_t}\big)^2=\big(\frac{\eta_{t}}{H_t}\big)^2\big(\frac{t_0+1}{t}\big)^2$, we have the equality
$ \big(\frac{\eta_{t_0}}{H_t}\big)^2=\frac{\alpha^2}{M^2}\big(\frac{t_0(t_0+1)}{t(t+1)}\big)^2$.
Hence, 
\begin{equation}\label{eq:sum_part1}
\sum_{i=1}^{t_0} \Big(\frac{\eta_{i}}{H_t}\Big)^2\leq \Big(\frac{t_0(t_0+1)}{t(t+1)}\Big)^2.
\end{equation}

By substituting \eqref{eq:sum_part2} and \eqref{eq:sum_part1} back into \eqref{eq:stepsplit}, we obtain
\begin{align*}
    \frac{\sqrt{\sum_{k=1}^t \eta_k^2}}{H_t}\!&\leq\! \sqrt{\Big(\frac{t_0(t_0+1)}{t(t+1)}\Big)^2\!+\!\Big(\frac{2}{t(t+1)}\Big)^2\sum_{k=t_0+1}^t k^2}%\\
   % & =: A(t,t_0),
\end{align*}
completing the   proof upon using the definition of $A(t,t_0)$.
\end{IEEEproof}
\end{appendices}

\bibliographystyle{IEEEtran}
\bibliography{reference}

\end{document}